\documentclass[useAMS,usenatbib]{mn2elettersize} 
\usepackage{amsmath}
\usepackage{natbib} 
\usepackage{graphicx}

\begin{document}

\pagerange{\pageref{firstpage}--\pageref{lastpage}} \pubyear{2003}
\def\lya{\mbox{Ly$\alpha$\phantom{ }}}
\def\lymana{\mbox Lyman $\alpha$\phantom{ }}
\def\ovititle{\mbox{O\,{VI }}}
\def\ovi{\mbox{O\,{\sc vi} }}
\def\civ{\mbox{C\,{\sc iv} }}
\def\ovimu{\mbox{${O\scriptstyle VI}$}}
\def\himu{\mbox{${H\scriptstyle I}$}}
\def\hi{\mbox{H\,{\sc i} }}
\def\la{\raisebox{-0.6ex}{$\,\stackrel
{\raisebox{-.2ex}{$\textstyle <$}}{\sim}\,$}}
\def\taurela{${\tau_{\rmn{O\textsc{VI},app}}-\tau_{\rmn{Ly}\alpha}}$\phantom{ }}
\def\overden{${n/\bar{n}}$\phantom{ }}
\def\overdencut{$(n/\bar{n})_{\rmn{cut}}$\phantom{ }}
\def\ga{\raisebox{-0.6ex}{$\,\stackrel{\raisebox{-.2ex}{$\textstyle >$}}{\sim}\,$}}
\def\la{\raisebox{-0.6ex}{$\,\stackrel{\raisebox{-.2ex}{$\textstyle <$}}{\sim}\,$}}

\def\mgh#1{{\bf MH:  #1}}
\def\mmp#1{{\bf MP:  #1}}
\def\cm{\,{\rm cm}}

\title[Pixel correlation searches for \ovititle in the \lymana forest]
    {Pixel correlation searches for \ovititle in the \lymana forest and
      the  volume filling factor of
    metals in the Intergalactic Medium at $z\sim 2-3.5$}
\author[Matthew M. Pieri and  Martin G. Haehnelt]
    {Matthew M. ~Pieri$^{1,3}$\thanks{Email: m.pieri@imperial.ac.uk} and Martin G. ~Haehnelt$^2$
    \\
    $^1$ Imperial College London, Blackett Laboratory, Prince Consort Road, London SW7 2BW, UK\\
    $^2$ Institute of Astronomy, Madingley Road, Cambridge CB3 0HA, UK\\
    $^3$ Max-Planck-Institut f\"ur Astrophysik,
    Karl-Schwarzschild-Str. 1, D-85740 Garching, Germany\\
    \\ }
\date{Accepted 2003 October 9. Received 2003 11 August}

\label{firstpage}

\maketitle
\begin{abstract}
Artificial absorption spectra are used to test  a  variety of
instrumental and physical effects on the the pixel correlation
technique for the detection of weak \ovi absorption. At \hi
optical depths $\la 0.3-1$, the apparent \ovi detections are
spurious coincidences due to  \hi absorption at other redshifts. In
this range, the {\it apparent} \ovi   optical depth is independent
of \hi optical depth. At larger \hi optical depths, apparent \ovi
optical depth and \hi optical depth are correlated. 
Detailed  modelling is required in order to interpret the significance 
of this relation. High-resolution spectra of four QSOs 
together with a large suit of synthetic spectra 
are used to show that the detection of \ovi in individual 
spectra is only statistically
significant for overdensities $\ga 5$. These overdensities are
larger than would be naively inferred from the onset of the
correlation and a tight optical depth-density relation.  
The lower limit for the volume filling factor of regions 
which are enriched by \ovi is 4\% at 95\% confidence.
This is no larger than the observed volume filling factor 
of the winds from Lyman break galaxies.  Previous claims that the
observed \ovi absorption extends to underdense regions and
requires a universal metal enrichment with large volume filling
factor, as may be expected from population III star formation at
very high redshift, appear not to be warranted.
\end{abstract}


\section{Introduction}

The \lya forest in QSO absorption spectra is now widely believed
to be caused by fluctuating  Gunn-Peterson absorption due
to an undulating warm photoionised Intergalactic Medium (IGM). With this
interpretation for the origin of the \lya forest, it has become
possible to relate the absorption optical depth to the density of absorption
 systems in a meaningful manner. According to this scheme low column density
absorption  systems arise from low density regions of the
Universe. A detection of associated metal absorption in weak
\lya absorption systems would suggest that the metal enrichment of the
IGM is widespread.

Models put forward to explain this metal enrichment fall into two
categories: they either propose that early 
and widespread Population III star formation in pre-galactic structures
at $z \ga 10$ is responsible (e.g. \citealt{1997MNRAS.291..505N}; 
\citealt*{2000MNRAS.319..539F}; \citealt{2001PhR...349..125B};
\citealt*{2001ApJ...555...92M}) or suggest a later episode of metal
enrichment by winds from starbursting galaxies at $z\le 5$ (e.g.
\citealt{2001ApJ...560..599A}, \citealt{2001MNRAS.321..450T},
\citealt{2003ApJ...584...45A}). Transporting metals from the
galaxies or protogalaxies where they are produced into the
(low density) IGM is not, however, a trivial matter. A variety of
mechanisms have been proposed.  Some have argued for
supernova-driven (\citealt{1986MNRAS.221...53C},
\citealt{1986ApJ...303...39D}) or `miniquasar'-driven
\citep*{1999ApJ...514..535H} winds. Others have argued for ejection
via galactic mergers (\citealt{1997ApJ...486..581G} and
\citealt{1998MNRAS.294..407G})  or photoevaporation during
reionisation \citep{1999ApJ...523...54B}. With an accurate
determination of the fraction of the universe 
(or `volume filling factor') enriched by metals it may be possible
to test the veracity of such models. A large volume filling factor
would favour early enrichment by stars forming in shallow 
potential wells, out of which metals are more easily
transported into low density regions.

Observationally, the most important tracers of metals in the
photoionised IGM are \civ and \ovi. Associated \civ absorption has
been detected down to  HI column densities of order
$10^{14}{\cm}^{-2}$ (\citealt{1995qal..conf..289T},
\citealt{1995AJ....109.1522C}, \citealt{2000AJ....120.1175E},
 \citealt{2003astro.ph..5413P})
corresponding to an optical depth of $1-10$. To investigate even lower
density systems a pixel correlation technique has been
employed where, instead of fitting an associated \civ absorption
line, a pixel-by-pixel search for excess absorption at the
corresponding \civ wavelength is performed
\citep{1998Natur.394...44C}. This method was hoped to be more robust
than the use of line `stacking'  (\citealt{1995qal..conf..289T}; 
\citealt{2000AJ....120.1175E}), but there is some ambiguity in  
the interpretation of the results. At 
the relevant redshifts, the \ovi species is the most sensitive
tracer of metals in low density regions of the IGM, provided that
the spectrum of the UV background is sufficiently hard
(\citealt*{1997ApJ...481..601R} and \citealt{1998ApJ...499..172H}).
However, the relevant  \ovi absorption lines
($\lambda_{O\textsc{vi}a} = 1032$\AA,
$\lambda_{O\textsc{vi}b} = 1038$\AA) occur in the same
wavelength range as the \lymana forest and higher order lines
of the Lyman series, which makes the detection of weak \ovi
absorption challenging.

\citet{2000ApJ...541L...1S} claim to have
detected \ovi at \hi optical depths significantly smaller than unity
in some QSO absorption spectra and used the optical depth-density
relation of numerical hydrodynamic simulations to argue that this
indicates a detection of metals in underdense regions of the IGM.
In this work, we apply the pixel-by-pixel search to
synthetic spectra and thereby investigate  instrumental and physical
effects on this technique (see also
\citealt*{2002ApJ...576....1A}). We then assess the statistical
significance of the \ovi detection and the implied density threshold
above which  the presence of metals is confidently detected.

The spectra (both simulated and observed) are described in Section
\ref{Simulated and Observed Spectra}.  In Section~\ref{Pixel-by-pixel search}
we outline the pixel correlation method used to detect the
presence of \ovi absorption. The results of the search
for \ovi are shown and discussed in Section
\ref{Results From Simulated Spectra}. In Section
\ref{Search for OVI in the low density IGM} we  
explore the consequences for the metal enrichment 
of the low density IGM.

\section{Simulated and Observed Spectra}
\label{Simulated and Observed Spectra}
\subsection{Calculating synthetic spectra}
\subsubsection{Basic Assumptions}

We use the method developed by Bi (\citealt*{1992A&A...266....1B};
\citealt{1997ApJ...479..523B})  to analytically
calculate synthetic spectra in the fluctuating Gunn-Peterson 
approximation for a warm photoionised IGM with  a density  distribution 
that is expected
in  a $\Lambda$CDM structure formation model. 
The basic
assumptions are:

\begin {enumerate}

\item the baryonic matter traces the  dark matter on scales
larger than a filtering scale, which is related to the Jeans scale;

\item the distribution of mildly non-linear
densities has a lognormal probability distribution function (PDF);

\item the absorbing gas obeys a simple temperature-density relation
      and is not shock-heated.
\end {enumerate}

Table~\ref{standardparam} shows the model  parameters. 
A brief summary of how  we calculate the synthetic spectra follows.

\begin{table}
 \caption{Model parameters used for all simulations}
 \label{standardparam}
 \begin{tabular}{@{}lc}
  \hline
  Parameter & Chosen Value \\
   \hline
     $\Omega_{\rm m}$ & 0.3 \\
     $h$ & 0.65 \\
     $\Omega_{\lambda}$ & 0.7 \\
     $\sigma_{\rm 8}$ & 0.9 \\
     $log[\bar T_{\rm {eff}}]$ & 4.6 \\
         $\mu$ & 0.641 \\
     $\Gamma$ & 1.333 \\
  \hline
 \end{tabular}
 \medskip

\end{table}

\subsubsection{The Spatial Distribution of Baryons}
\label{The Spatial Distribution of Baryons}
According to standard
structure formation models the matter distribution was initially a
Gaussian random field with small density fluctuations that evolve
under gravity into the pattern of sheets and filaments
characteristic of CDM models. Many features of such a distribution
can be reproduced by a rank-ordered mapping of a linear Gaussian
random field onto a lognormal PDF of the density. This idea is
at the heart of the method used to calculate synthetic absorption
spectra \citep{1992A&A...266....1B}.  This method is an 
efficient way of producing large numbers
of realistic synthetic spectra.
Note that we are interested in producing
synthetic spectra with a realistic density PDF and realistic
instrumental properties. Reproducing
the detailed clustering properties of the density field  is less 
important.

We begin by creating a linear Gaussian random field which is fully
determined by its power spectrum, $P_{DM}(k)$, for which 
we have used the form given by
\citet*{1992MNRAS.258P...1E},

\begin{equation} \label{powspec}
P_{\rm DM}(k)\propto \frac{k} { (1+ [ak+(bk)^{3/2}+(ck)^2]^\nu )^{2/\nu} },
\end{equation}

where $a=6.4/\Gamma$, $b=3.0/\Gamma$, $c=1.7/\Gamma$,
$\Gamma=\Omega_mh$ ($\Omega_m$ \& $h$ have
 their normal definitions) and $\nu=1.13$.
The power spectrum was normalised  by the {\it rms} fluctuation
amplitude on a $\mathrm{8h^{-1}Mpc}$ scale,  $\sigma_8$,  as given in 
Table 1. In order
to take into account pressure effects that suppress fluctuations
on scales smaller than  a certain filtering scale
\citep{1998MNRAS.296...44G} the power spectrum of the baryon
density in the linear regime is assumed to have the form,

\begin{equation} \label{jeans}
P_{\rm B}(k)=\frac{P_{\rm{DM}}(k)}{(1+x_{\rm b}^2k^2)^2},  \qquad
x_{\rmn{b}}=\left[\frac{2\gamma k \bar{T}_{\rm{eff}}} {3\mu
m_{\rm p}\Omega(1+z)}\right]^{\frac{1}{2}},
\end{equation}

where $x_{\rmn{b}}$ is a comoving scale related to the Jeans length
(\citealt{1980lssu.book.....P}; \citealt{1997ApJ...479..523B}),
 $\gamma$ is the ratio of specific
heats, $\bar{T}_{\rm{eff}}$ is the effective mean temperature and $\mu$ is
the molecular weight of the IGM. An appropriate value for
$\bar{T}_{\rm{eff}}$ has been determined from comparison with
hydrodynamic simulations by \citet{1997ApJ...479..523B}.
This is the baryonic matter power spectrum at $z=0$
and linear theory is used to  evolve it backward to high redshift.

Equations (\ref{powspec}) and (\ref{jeans}) provide the 3D power spectrum.
A 1D power spectrum needed for the description of
line-of-sight (LOS) fluctuations is obtained by an integration of
the 3D power spectrum \citep{1991ApJ...379..482K}.
The  matter density  and the peculiar velocity
field are related and can be described as coupled
Gaussian random fields  (see \citealt{1992A&A...266....1B} for more details).
Random realisations of the density  and
peculiar  velocity in real space (co-moving coordinates) are
obtained using a Fourier transform routine.  The non-linear evolution
is modelled by a local rank-ordered mapping to a lognormal
density probability distribution,

\begin{equation} \label{lognorm}
n/ \bar n =
\exp\left[\delta(x)-\frac{\langle \delta^2 \rangle}{2}\right].
\end {equation}

where $\delta = (n -\bar n)/\bar n$ and $n/ \bar n$ is the overdensity
factor. At the relevant redshift the baryonic density field is still
in the mildly non-linear regime and the use of a log-normal PDF is
a reasonable assumption (\citealt{1992A&A...266....1B};
\citealt{2000MNRAS.313..364N}).

\subsubsection{From the density to the absorption spectrum }
\label{General Simulation of Spectra}

The \hi optical depth depends not only on the density but
also on the temperature of the gas, due to the temperature dependence
of the  recombination rate. For densities $n/\bar n \la10$,
numerical simulations show that most of the gas is not shocked and
that a simple power law relation between density and temperature
is established by the balance of photoionisation heating and
adiabatic cooling \citep{1997MNRAS.292...27H},

\begin{equation}
\label{tempden}
 T=T_0(n/\bar n)^\alpha,
\end {equation}

where $0.3\la\alpha\la0.6$  and $T_0$ is a constant. 

For  gas at  temperatures of a few times $10^4K$ 
in photoionisation equilibrium the neutral
hydrogen density is proportional to
$n^2T^{-0.7}/\Gamma_{\mbox{H\,{\sc i}}}$, where $\Gamma_{\hi}$ is the
photoionisation rate. As a result, the optical depth to \lya 
absorption at redshift $z$ can be written 
as

\begin{table}

 \caption{Observed Data Sample}
 \label{datasample}
 \begin{tabular}{@{}lcccc}
  \hline
      QSO & $\mathrm{z_{em}}$ & $\mathrm{z_{range}}$
      & $\mathrm{\bar{D}_{\textsc{ovi}}}$
      &  $\mathrm{\bar{D}_{\textsc{hi}}}$\\
  \hline
     Q1122-165 & 2.40 & 2.02-2.34 & 0.157 & 0.165\\
     Q1442+293 & 2.67 & 2.51-2.63 & 0.194 & 0.193\\
     Q1107+485 & 3.00 & 2.71-2.95 & 0.254 & 0.253\\
     Q1422+231 & 3.62 & 3.22-3.53 & 0.418 & 0.386\\
  \hline
 \end{tabular}
 \medskip
\end{table}

\begin {equation}
\label{optdep}
\tau_{\rmn {Ly}\alpha}(z) =  \sigma_0 \bar n_{\rm HI} I (z), 
\end {equation}

where

\begin {eqnarray}
\label{integral}
\lefteqn{I (z)= \int_{z_1}^{z_2} \frac{c dz^\prime}{1+z} \frac{c}{H_0 E(z^\prime)} \left(
\frac {n(z^\prime)} {\bar n} \right)^{1.7} }
                    \nonumber\\
  & &  {} \frac{1}{b} \exp \left(-\left(\frac {c( z-z^\prime)} {
  (1+ z)}+v_{\rm pec}(z^\prime) \right) ^2 \Bigg/ b^2\right),
\end {eqnarray}

$z_1$ and $z_2$ are the lower and upper limits 
of the redshift region of interest, $b$ is the
Doppler parameter, $n_{\rm HI}$ is the neutral hydrogen density,
$v_{\rm pec} $ is the peculiar velocity and a value  $\alpha=0.4$ 
is assumed. $E(z)$ is the redshift dependent component of the Hubble parameter 
and $\sigma_0$ is the cross-section for  resonant \lya scattering.

The effect of peculiar
velocity and Doppler broadening are taken into account to obtain
the optical depth in redshift space.   We have thereby
approximated the Voigt profile  by a Gaussian distribution, which
is a reasonable approximation for  absorption lines with equivalent width
$<0.7$\AA.

The optical depth for the \ovi doublet is calculated
 assuming a fixed  ratio of
\ovi number density to \hi number density
($n_{\rmn{O\textsc{VI}}}/n_{\rmn{H\textsc{I}}}$). The optical depth of the
higher order Lyman lines are also calculated.

\subsubsection{Realistic Synthetic Spectra}
\label{Realistic Synthetic Spectra}

To facilitate a more accurate comparison with the observed spectra
we model the optical depth distribution for the same wavelength
range as the observed spectra. There are two regions of the
spectra of particular interest; the `\ovi region', where \ovi
absorption is searched for, and the `\hi region', where the
corresponding \lya absorption  occurs. The spectra are obtained
by co-adding the optical depth of  \mbox{Ly$\alpha$}, \ovi and higher order
Lyman series lines. We then use the estimated statistical error on
the flux for the observed
spectrum to add random noise and perform a Gaussian smoothing to
mimic the effect of instrumental broadening.

The optical depth is scaled such that the mean flux decrement
of the observed  QSO is reproduced.  As we will discuss later the
results of the pixel-by-pixel search depend sensitively on the
mean flux level. We have thus rescaled the optical depth for the
\hi and the \ovi regions independently to reproduce the observed
mean  flux level in both regions ($\bar D_{\rmn {\textsc{hi}}}$
and  $\bar D_{\rmn {\textsc{ovi}}}$ respectively). $\bar
D_{\rmn{\textsc{hi}}}$ is used to set $\bar{\tau}_{\rmn{Ly}\alpha}$
in the \hi region as well as $\bar{\tau}_{\rmn{O\textsc{vi}}}$ 
(with fixed $n_{\rmn {O\textsc{VI}}}/n_{\rmn {H\textsc{I}}}$) and 
the optical depth  of the higher order Lyman lines
in the \ovi region.  $\bar D_{\rmn{O\textsc{vi}}}$ is then used to
set $\bar{\tau}_{\rmn{Ly\alpha}}$ in the \ovi region. Note that
this means that we allow a difference in the
value of $\Omega_{\rmn{ bar}}^{2}/\Gamma_{\rmn {H\textsc{i}}}$ in the two
regions of the spectrum. It seems plausible to assume that the
errors we obtain in this way for a range of Monte Carlo
realisations of the LOS density distribution are similar to those
we would have found if we had only picked random realisations
that can reproduce the mean flux decrement in the \hi and the
\ovi regions simultaneously. The latter approach is 
computationally prohibitive.

In this way we  have  obtained ensembles of simulated spectra with
the same wavelength range, mean flux decrement  and noise
properties as the observed  spectra, but differing in their
specific random realisation of LOS matter distribution and
noise.

\subsection{Observations} \label{Observations}

We have used four observed QSOs which are part of the sample used by
\citet{2000ApJ...541L...1S} for a detailed comparison with synthetic QSOs spectra:
Q1122-165,  Q1442+293, Q1107+485 and Q1422+231. The data for Q1122-165
was taken during Commissioning I and Science Verification
observations for the UVES instrument on the VLT (Kueyen) and
released by ESO for public use. The reduction method can be found
in \citet*{2001A&A...373..757K}. The other spectra were taken with
the HIRES instrument \citep{1994SPIE.2198..362V} on Keck I and
reduced using procedures described in \citet{1997AJ....113..136B}.
These two instruments provide spectra with resolutions
$R_{\rmn{max}}\approx$ 80000 and 110000 respectively and have typical
S/N of 50.  Table \ref{datasample} gives a list of the redshift range and 
flux decrement for the \ovi and \hi regions. Regions of strong absorption 
from other known metal lines were removed.

\section{Pixel-by-pixel search}
\label{Pixel-by-pixel search}

\begin{figure}
\includegraphics[angle=90,width=.95\columnwidth]{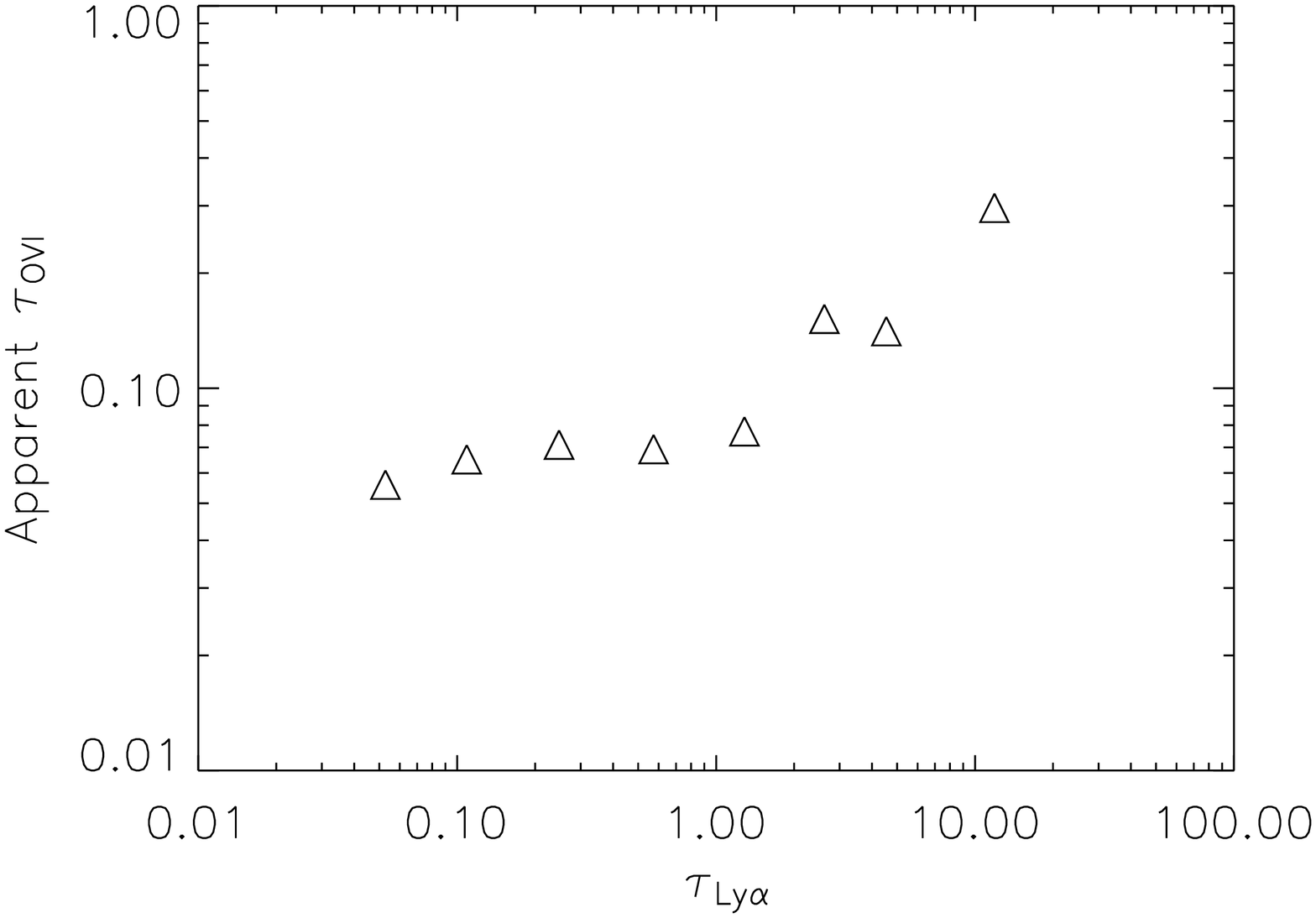}
\caption{An example of the results from the pixel-by-pixel search
method for the absorption spectrum of Q1107+485.} 
\label{Search1107}
\end{figure}

We use a technique for the detection of \ovi in the \lymana forest
initially developed by \citet{1998Natur.394...44C}
for the detection of weak \civ absorption.
\citet{2000ApJ...541L...1S} adapted this method to search for the
\ovi doublet (see also \citealt{2002ApJ...576....1A}). 
We briefly summarise the  procedure here.
The redshift range explored in each of the four 
QSOs is shown in Table \ref{datasample}. This is the same as used by 
\citet{2000ApJ...541L...1S}. Due to the increase in noise we did not find 
it worthwhile to extend the search to lower wavelength.

The continuum-fitted absorption spectrum is re-binned to 
pixels of equal size in redshift space for the Lyman series and
both members of the \ovi doublet.
A width of
$(c\Delta\lambda/\lambda_{\rmn {Ly}n}\rmn{)kms^{-1}}$ is taken (where 
Ly{\it n} is the highest order Lyman line used). In order to take into
account possible velocity shifts between \ovi and \hi absorption, 
we have also tried larger pixels widths. This did not improve the 
sensitivity for \ovi detection.

A search for correlated absorption at the wavelengths of the Lyman 
series and the \ovi doublet is then performed. Each pixel of the \lya 
absorption region is considered in turn.
 If the  flux level is within
$\sigma_{\rmn {noi}}/2$ of the continuum level (where $\sigma_{\rmn {noi}}$ is the estimated 
statistical error on the flux for the observed spectrum) or in a region where the
associated \ovi absorption is not covered the pixel is discarded.
If the pixel is within $\sigma_{\rmn {noi}}/2$ of saturation we attempt 
to use the higher order Lyman lines to estimate 
$\tau_{\rm{Ly}\alpha}$. We thereby use the higher order line with the lowest
equivalent \lya optical depth $\tau_{{\rm Ly}\alpha}=\min
(\tau_{{\rm Ly}n}f_{{\rm Ly}\alpha}/f_{{\rm Ly}n}\lambda_{{\rm Ly}n})$ that is not within
$\sigma_{\rmn{noi}}/2$ of saturation or the continuum (where $f_{{\rm 
Ly}n}$ is the
oscillator strength of the  Ly{\it n} line).

Once we have established a value for $\tau_{\rmn{Ly}\alpha}$, we consider
the pixels corresponding to absorption by the \ovi doublet at
the same redshift. The limiting factor in the use of this method
is coincident absorption from \mbox{H\,{\sc i}}, either in the form of higher
order Lyman lines at similar redshift or by \lya lines at lower
redshift. Obviously we want to minimise this
contamination. The doublet nature of the \ovi is utilised in this
regard. The smaller equivalent absorption,

\begin {equation}
\label{tauovimin}
\tau_{\rmn{O\textsc{vi}}} = {\rm min} 
\left(\tau_{\rmn{O\textsc{vi}a}},\frac{f_{\rmn{O\textsc{vi}a}}
\lambda_{\rmn{O\textsc{vi}a}}\tau_{\rmn{O\textsc{vi}b}} }
{f_{\rmn{O\textsc{vi}b}}\lambda_{\rmn{O\textsc{vi}b}} }\right),
\end {equation}

will be the least contaminated. This gives us
an upper limit to the \ovi optical depth. We will thus 
call the \ovi optical depth obtained in this way the
apparent \ovi optical depth. 

As pointed out  by \citet{2002ApJ...576....1A} the contamination
by higher order  Lyman lines  can be estimated from the
corresponding \lya absorption and corrected. Such contamination is subtracted from the \ovi optical depth where possible. If this is not
possible because of saturated \lya absorption we also discard the
pixel.

We then bin the ensemble of optical depth pairs in  \lya optical depth
and plot the relation between the median optical
depth of \ovi and \lya. We use the median instead of the 
mean as the distribution
of \ovi optical depth in each bin is skewed toward high optical
depths due to the contamination by \hi absorption.

Fig.~\ref{Search1107} shows an example of the relation of the
apparent \ovi and \lya optical depth (\taurela relation) obtained
in this way. At small \lya optical depth, the median \ovi optical
depth is constant. As we will demonstrate in detail in the next
section (see also \citet{2002ApJ...576....1A}), the apparent \ovi is here 
almost exclusively due to contamination of coincident \hi absorption 
and this region of the diagram {\it cannot} be used to infer a 
detection of \ovi. At larger
\lya optical depth, the apparent \ovi optical depth rises and is
mainly due to the presence of \ovi.  The level of metal enrichment is,
however, not straightforward and requires detailed modelling.

\begin{figure}
\centering
\includegraphics[angle=0,width=.85\columnwidth]{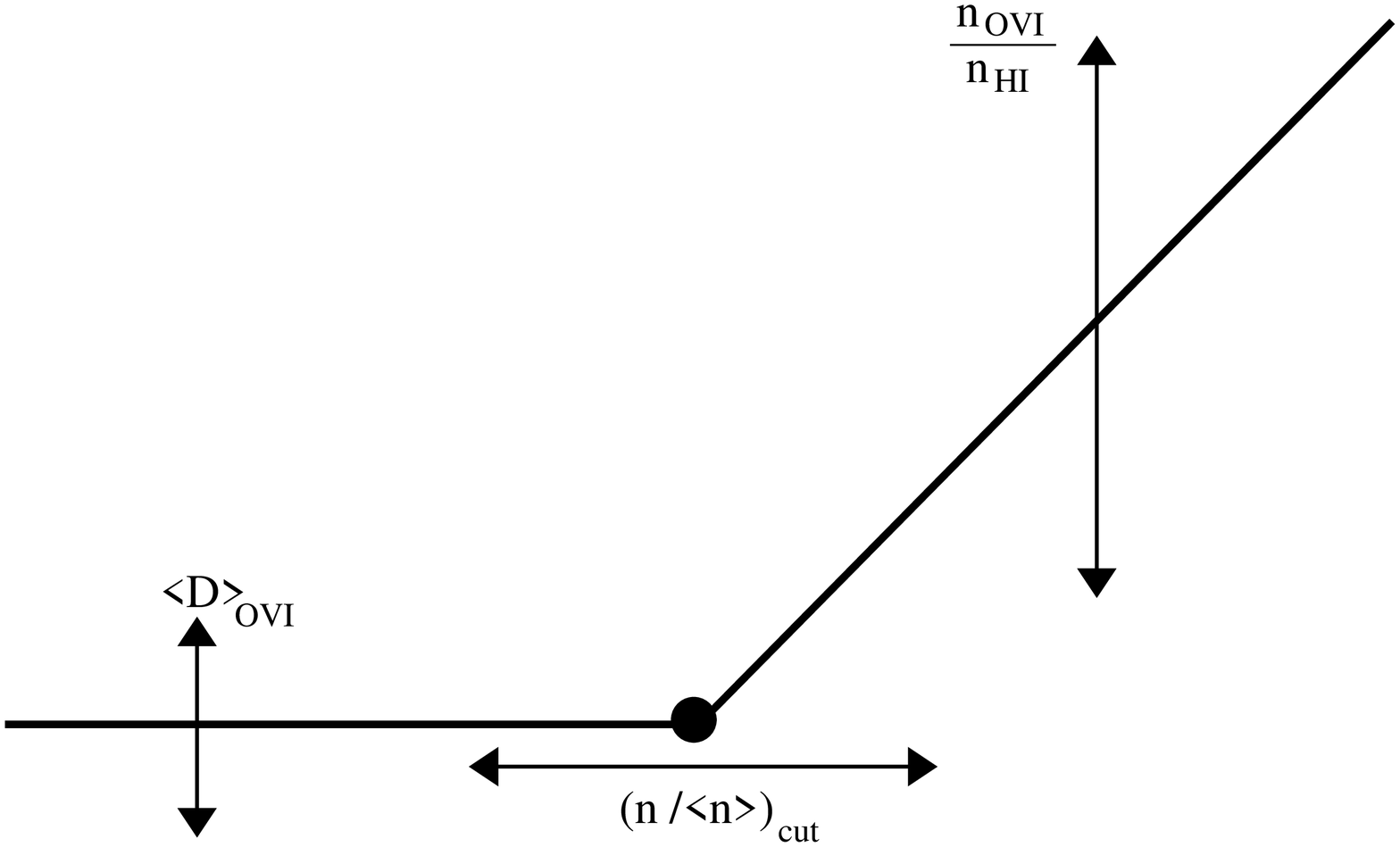}
\caption{A sketch showing the dependence of the \taurela relation 
on the principal  parameters $\bar D_{\rmn{O\textsc{vi}}}$, 
$n_{\rmn{O\textsc{VI}}}/n_{\rmn{H\textsc{I}}}$ and $(n/\bar{n})_{\rmn{cut}}$. }
\label{ovisearchsketch}
\end{figure}

\begin{figure*}
\includegraphics[angle=90,width=.999\columnwidth]{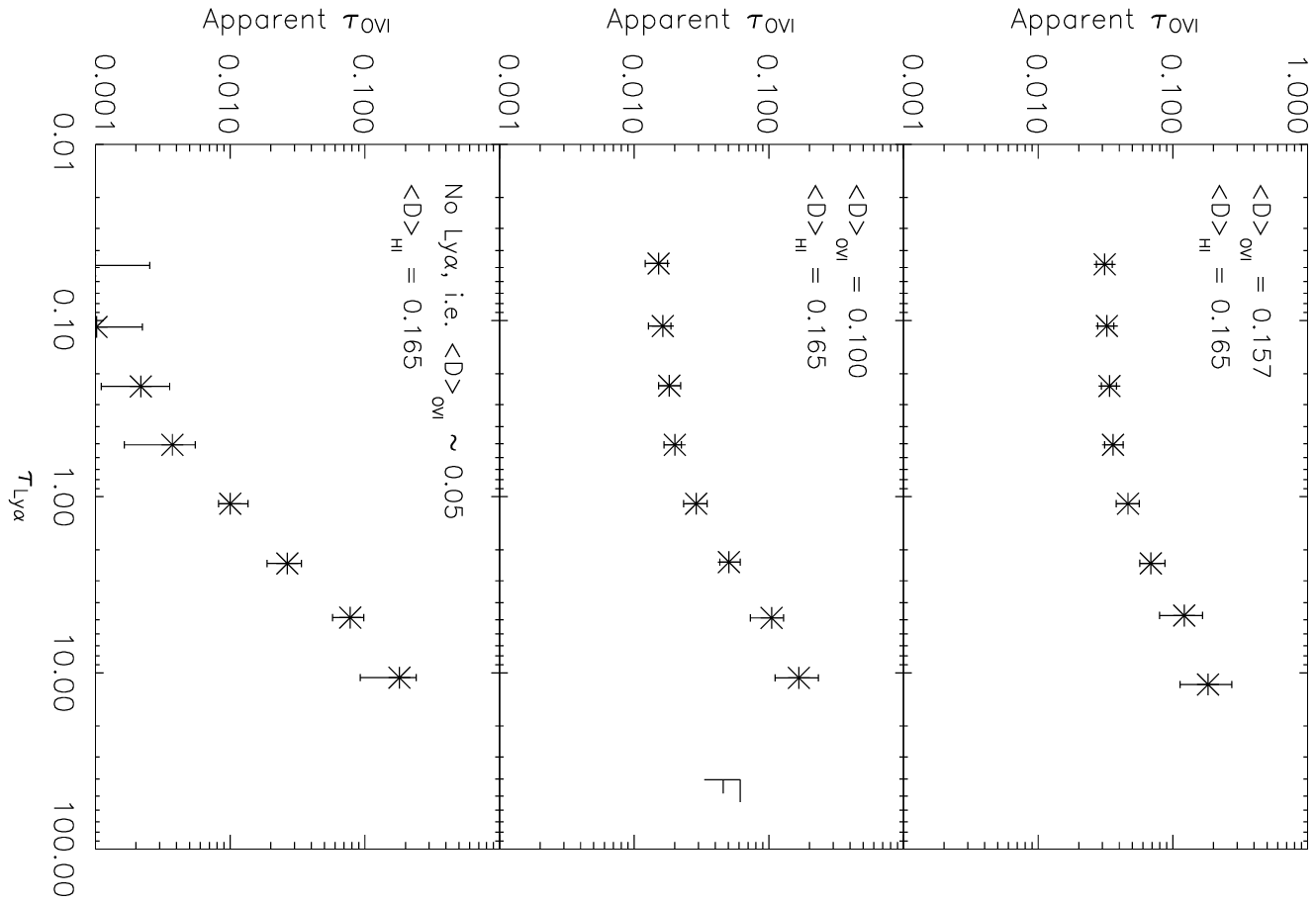}
\includegraphics[angle=90,width=.999\columnwidth]{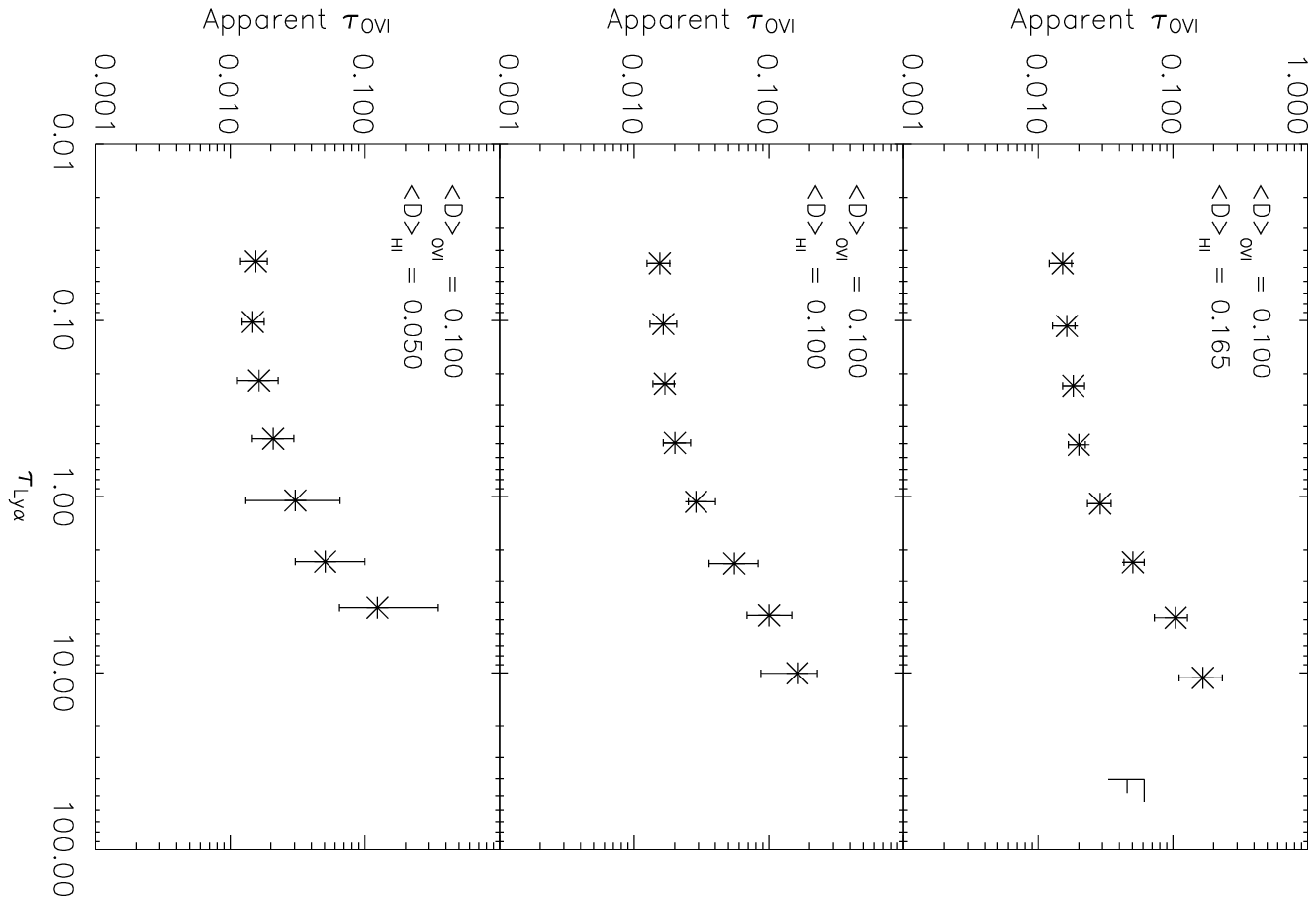}
\caption{The \taurela relation for an ensemble of forty synthetic spectra. 
{\it Left:} $\mathrm{\bar D_{O\textsc{VI}}}$ increases from bottom to
top. {\it Right:} $\mathrm{\bar D_{H\textsc{I}}}$ increases from bottom to top}
\label{simmfdvary}
\end{figure*}

\section{Results From Simulated Spectra}
\label{Results From Simulated Spectra}

\subsection{Principle parameters}

To obtain an indication of how the pixel-by-pixel search depends
upon instrumental effects and physical parameters and what errors
should be expected  due to variation of the LOS density
distributions, we investigated a large suit of synthetic spectra.
We have varied  the mean flux decrement in the \ovi and \hi
regions ($\bar D_{\rmn{O\textsc{vi}}}$ and $\bar D_{\rmn{H\textsc{i}}}$),
 the ratio  of \ovi to \hi density
($n_{\rmn{O\textsc{vi}}}/n_{\rmn{H\textsc{i}}}$) and introduced a varying 
cut-off in the density below which we set the \ovi density to zero 
(\overdencut).
Fig.~\ref{ovisearchsketch} sketches how 
the relation of apparent \ovi optical depth and \lya optical depth
depends on the three key parameters 
$\bar D_{\rmn{O\textsc{vi}}}$, \overdencut and 
$n_{\rmn{O\textsc{vi}}}/n_{\rmn{H\textsc{i}}}$.
A detailed discussion of the effect of these and the other parameters 
using synthetic spectra of Q1122 will  follow below. 
The parameters were varied around fiducial values 
denoted by an `F' in  figures~\ref{simmfdvary}-\ref{simnoisecontinvary}.  
The error bars in these figures are the $1\sigma$ spread of the
optical depths for ensembles of 40 spectra.

\subsection{Varying the mean optical depth}

Fig.~\ref{simmfdvary} shows how changing the mean
flux decrement affects the \taurela relation. In the left panels  we  consider 
changes in the \ovi region, $\bar D_{\rmn{O\textsc{vi}}}$. 
With increasing flux decrement the level of spurious coincidences 
due to \lya absorption from gas at a lower redshift rises 
and mimics the detection of \ovi. Since this \lya absorption is 
uncorrelated with absorption in the \hi region, this leads to a 
floor of constant apparent \ovi optical depth which rises with increasing
$\bar D_{\rmn{O\textsc{vi}}}$. In the right panels of 
Fig.~\ref{simmfdvary} we vary the
mean flux decrement in the \hi region $\bar D_{\rmn{H\textsc{i}}}$.
This time the shape of the relation between \hi optical depth and
apparent \ovi optical depth does not change but the error bars
increase substantially with decreasing mean flux decrement. This
is because, for small mean flux decrement, large optical depth \lya
pixels are poorly sampled.

\subsection{Varying the \ovi distribution}

\begin{figure*}
\centering
\includegraphics[angle=90,width=.999\columnwidth]{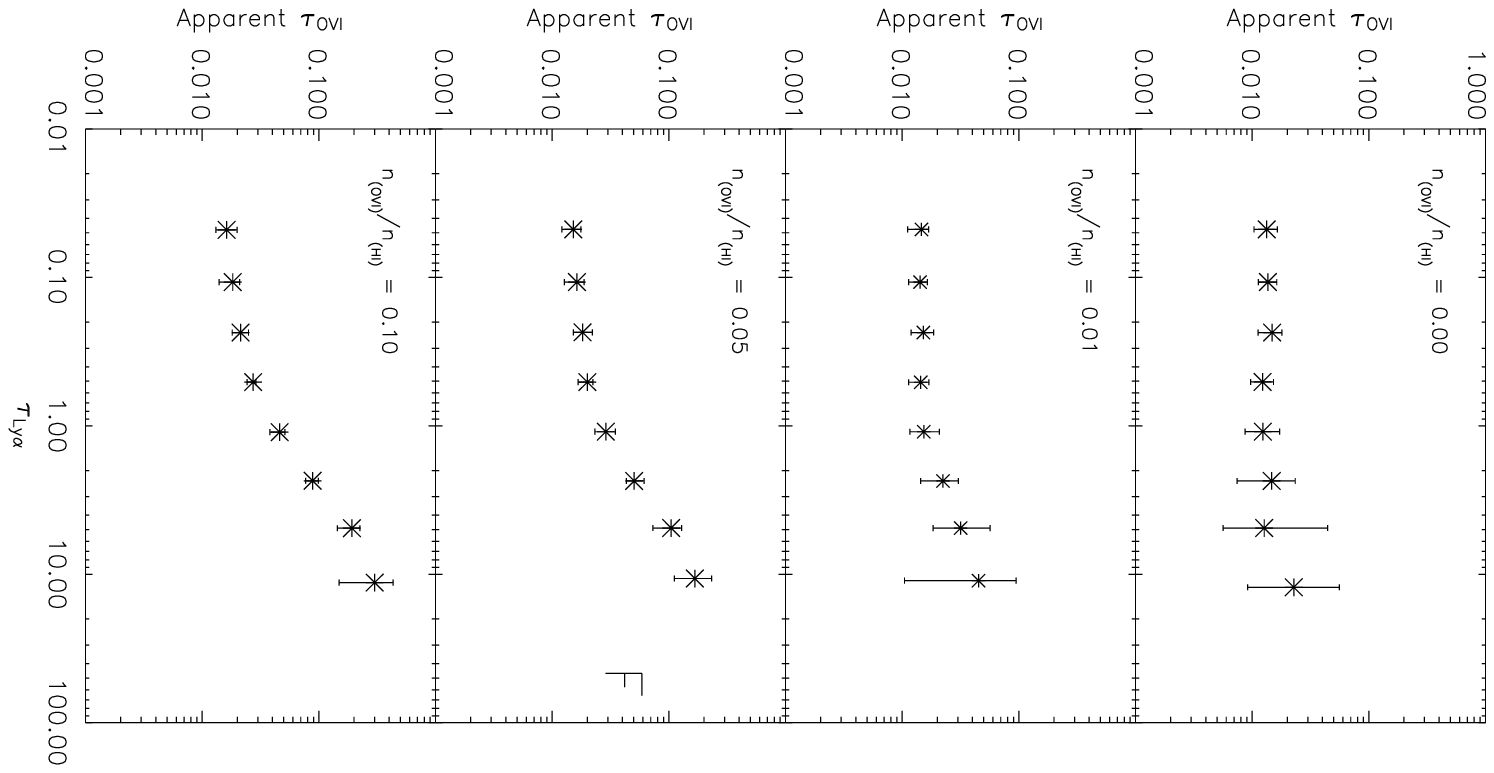}
\includegraphics[angle=90,width=.999\columnwidth]{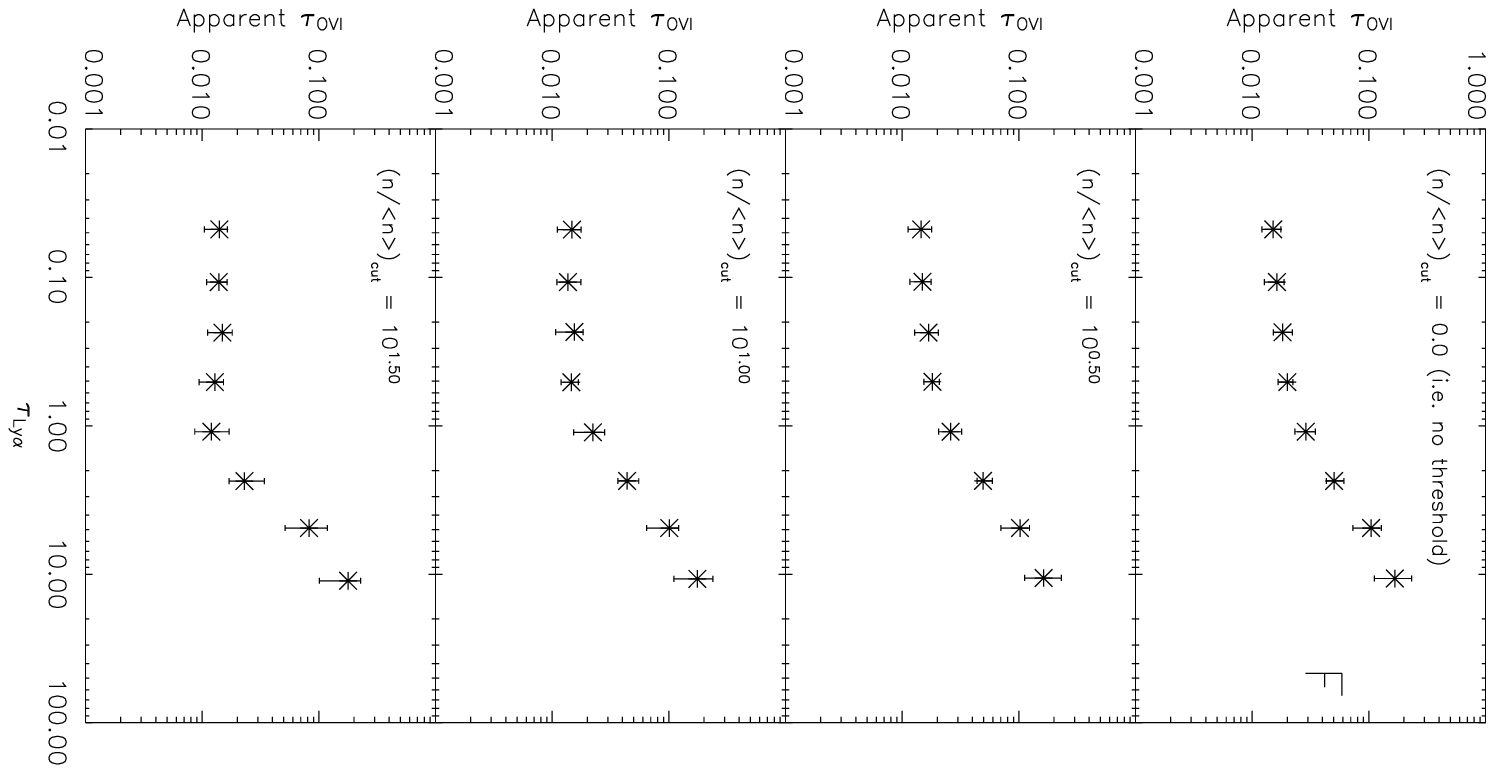}
\caption{As Figure \ref{simmfdvary}. {\it Left:} $n_{O\textsc{vi}}/n_{H\textsc{i}}$ 
increases from top to bottom. {\it Right:} \overdencut increases from top to bottom.}
\label{simovidistvary}
\end{figure*}

In Fig.~\ref{simovidistvary} we vary the \ovi distribution for our 
synthetic spectra. The left panels shows
how  varying the ratio of \ovi to \hi density
affects the  \taurela relation. With 
increasing ratio a correlation between apparent \ovi
optical depth and \hi optical depth develops for large \hi optical
depth but the errors are large.  As discussed in the introduction
the main aim of the search for \ovi at small \hi optical depth is
to establish the volume filling factor of metals in the Universe.
Because of the possible spurious detections due to random coincidences this
is not straightforward. 

There is another important effect which
can lead to spurious detection of \ovi. Low \lya optical depth is not
necessarily associated with low density gas but can also be due
to the wing of a strong \lya absorption feature which is caused by
a high density region. In order to test the density range for
which the presence of \ovi is required to reproduce the observed
\taurela relation, we introduce a cut-off in the density,
\overdencut below which we set the \ovi density to zero, i.e.

\begin{equation}
\label{delcutrangezero}
\frac {n_{OVI}} {n_{HI}}  =  0,  \quad \textrm{if} \quad
\frac {n} {\bar n}  <  \left(\frac{n} {\bar{n}}\right)_{\rmn{cut}},
\end{equation}

\begin{equation}
\label{delcutrangeconst}
\frac { n_{OVI} } {n_{HI}}=\textrm{const}, \quad \textrm{if} \quad
\frac {n} {\bar n}  \ge  \left(\frac{n} {\bar{n}}\right)_{\rmn{cut}}.
\end {equation}

In the right panel of Fig.~\ref{simovidistvary} we vary 
$(n/\bar{n})_{\rmn{cut}}$. The onset of the
correlation   moves to larger \hi optical depth.
Unfortunately, this dependence is  weak.

\subsection{Varying S/N and continuum level}

\begin{figure*}
\centering
\includegraphics[angle=90,width=.999\columnwidth]{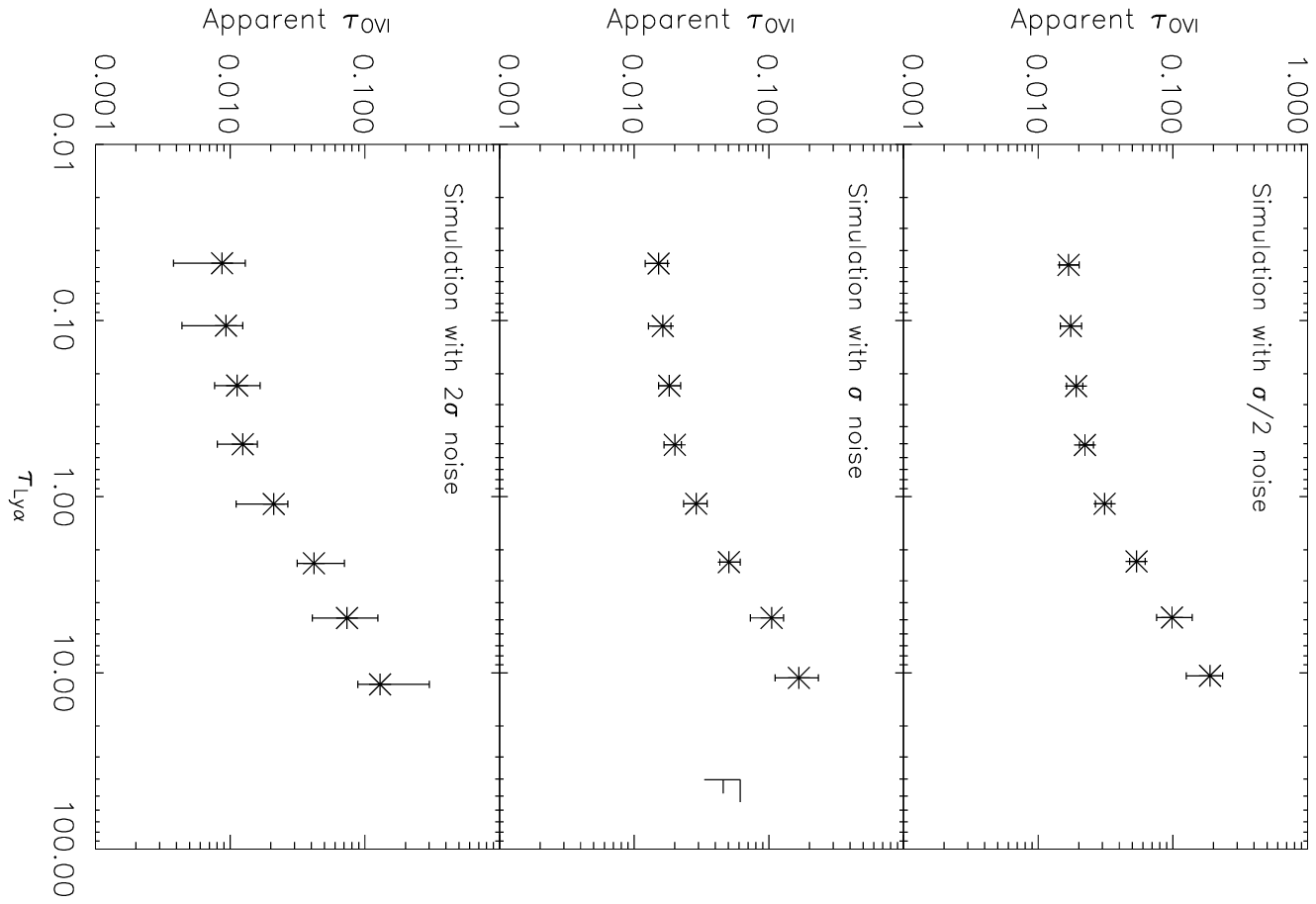}
\includegraphics[angle=90,width=.999\columnwidth]{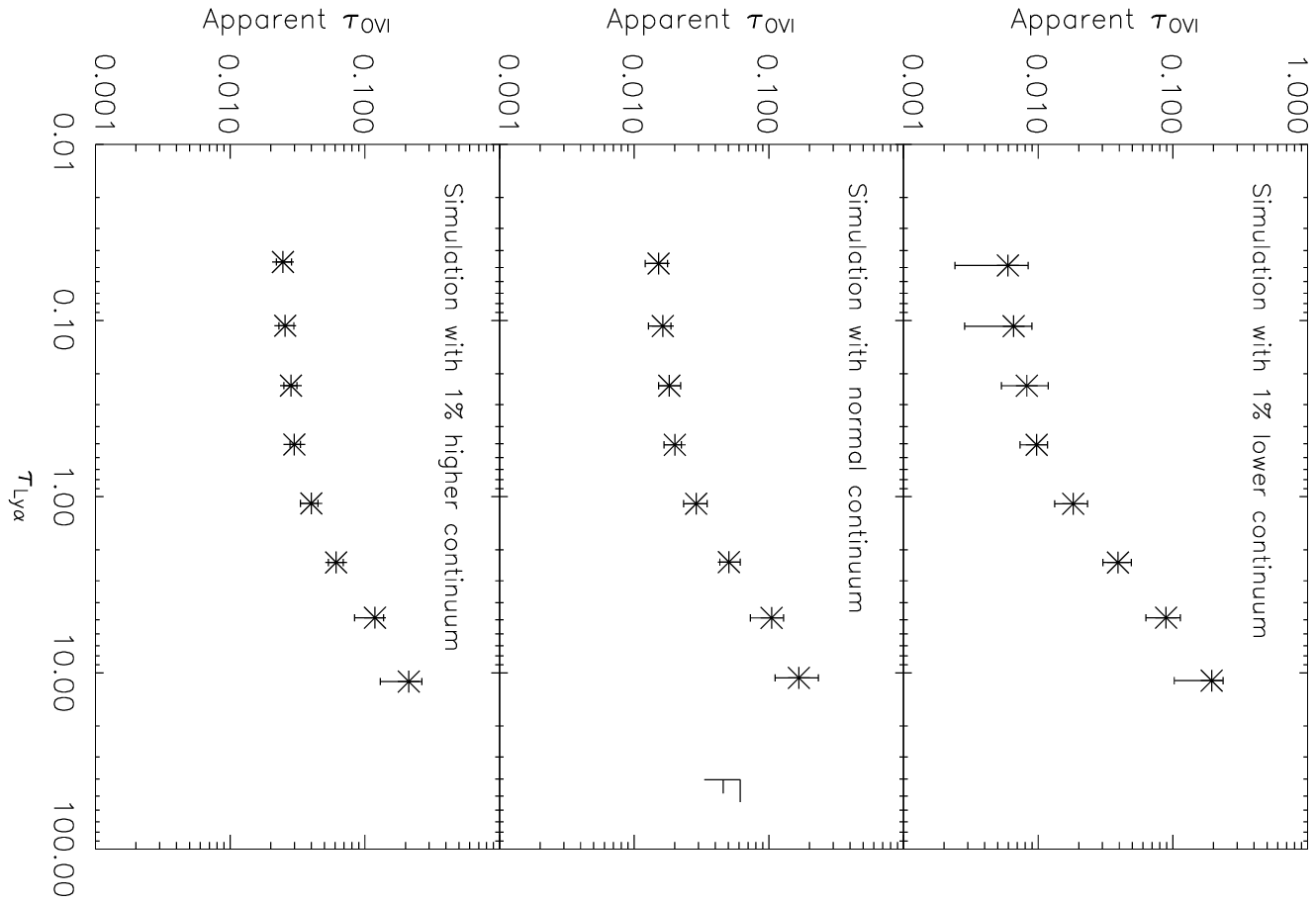}
\caption{As Figure \ref{simmfdvary}. {\it Left:} noise increases from top to 
bottom.  {\it Right:} the  continuum level increases from top to bottom.}
\label{simnoisecontinvary}
\end{figure*}

The left panels of Fig.~\ref{simnoisecontinvary} 
show the effect of changing the simulated noise. 
As expected, the errors increase significantly with 
increasing  noise level. 

The effect of an error in the continuum fit is shown
in the right panels of Fig. \ref{simnoisecontinvary}, 
where we have raised and lowered the 
continuum as a whole by 1\% in each synthetic spectrum.  
When we raise the continuum level the apparent \ovi optical depth 
increases because the optical depth of  spuriously coincident 
\hi absorption increases.  Note that if the continuum is placed too low
the  correlation between \hi and \ovi optical depth misleadingly 
appears to extend  to smaller \hi optical depth.

\begin{figure*}
\centering
\includegraphics[angle=90,width=.89\columnwidth]{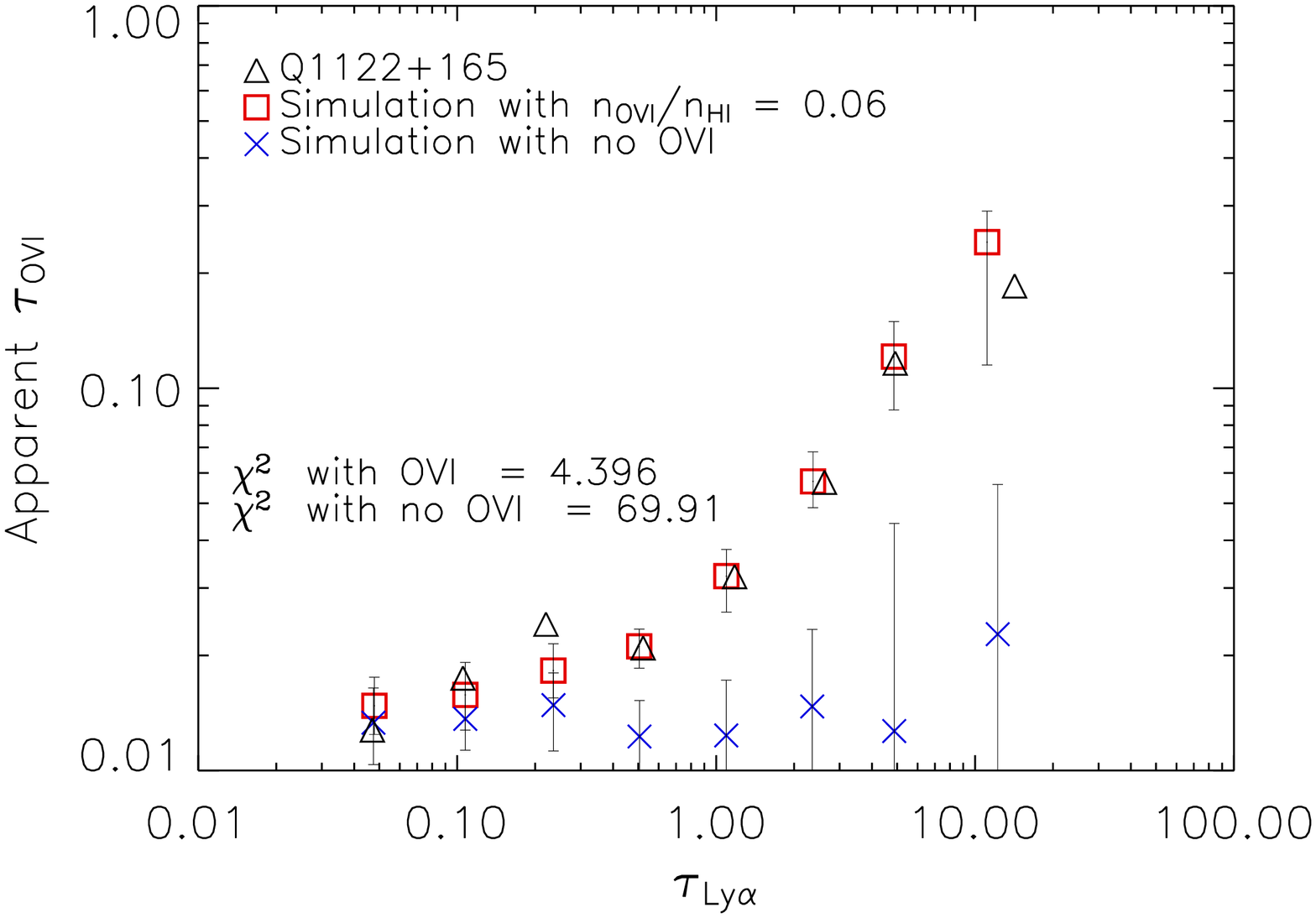}
\includegraphics[angle=90,width=.89\columnwidth]{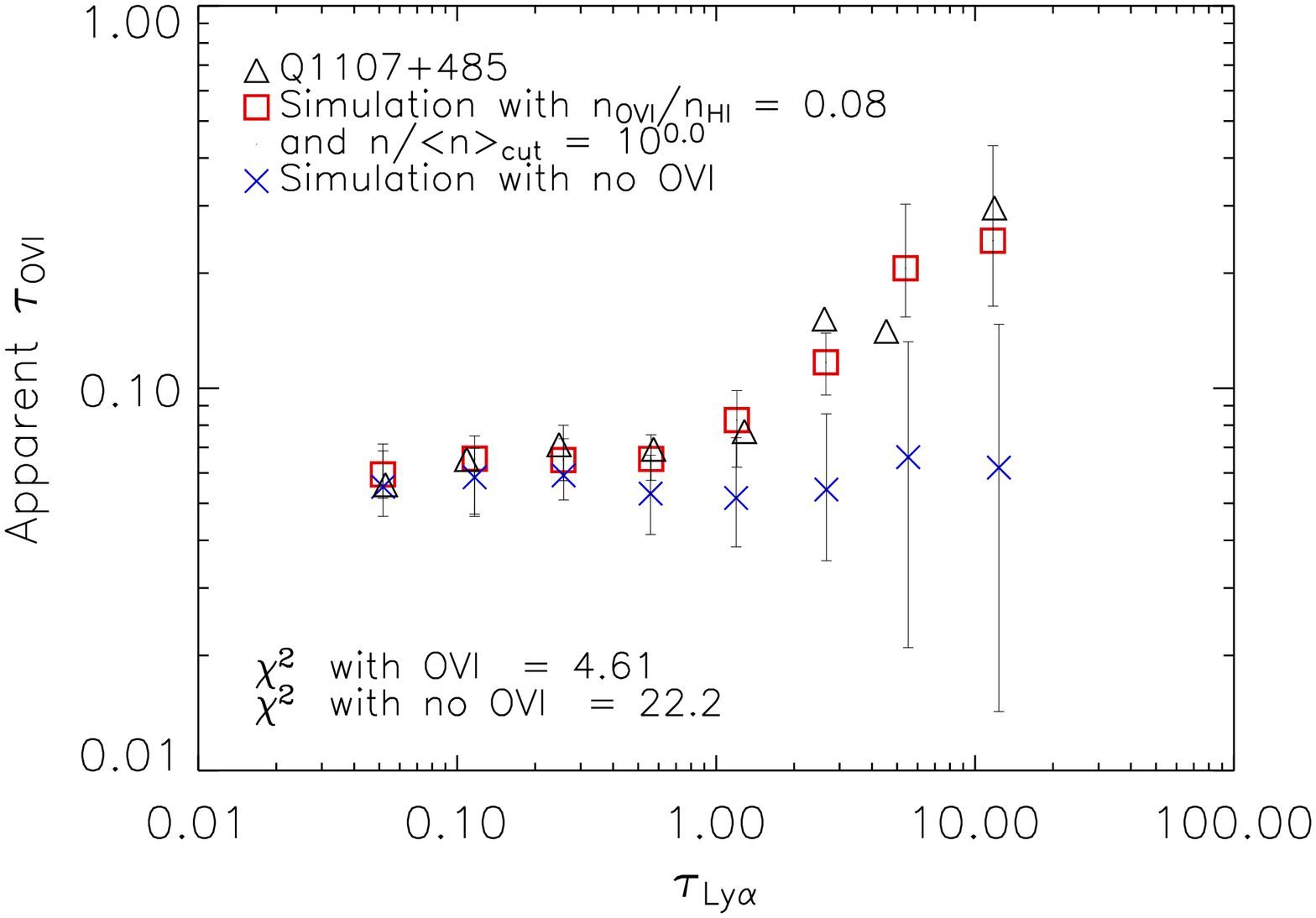}
\includegraphics[angle=90,width=.89\columnwidth]{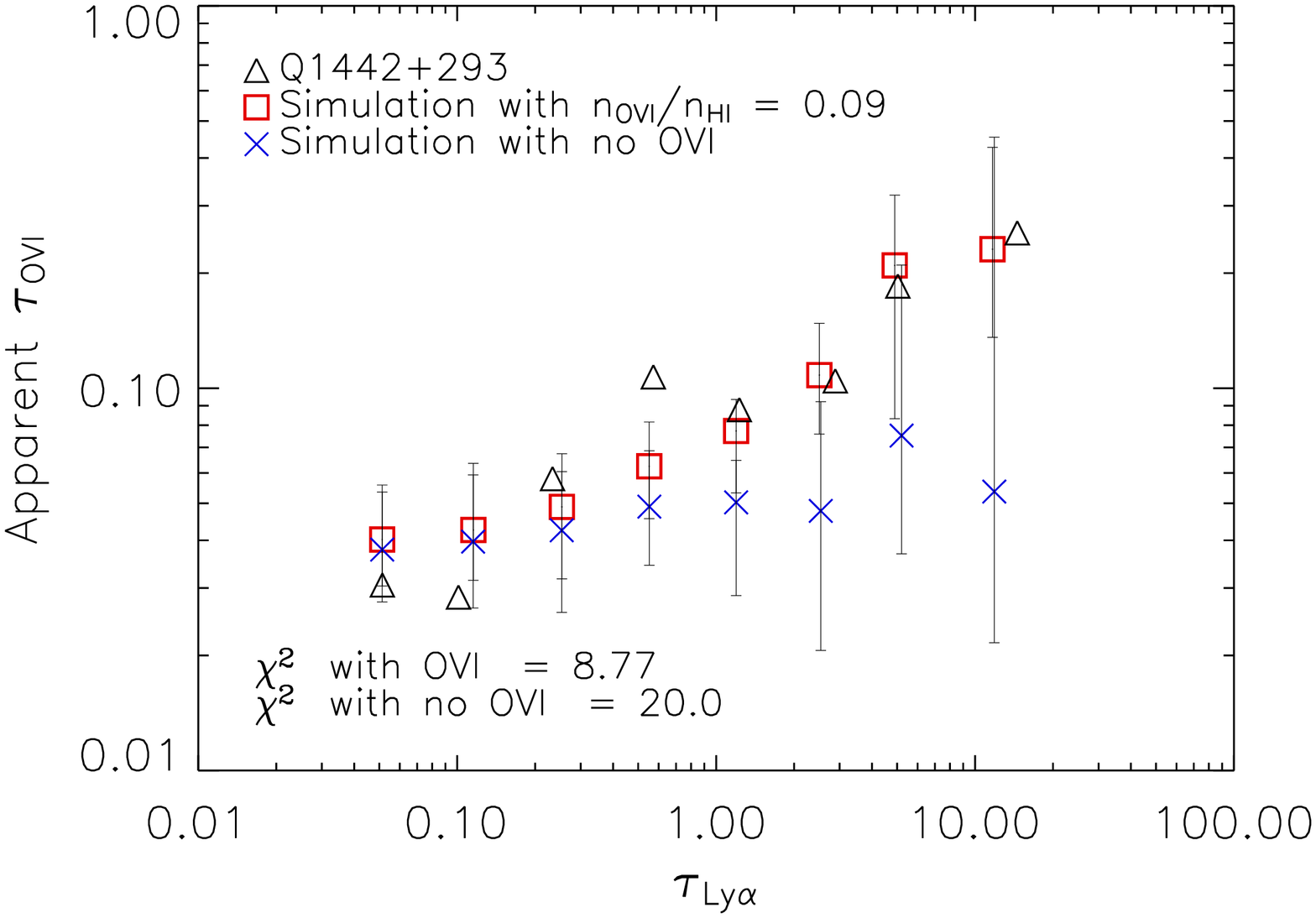}
\includegraphics[angle=90,width=.89\columnwidth]{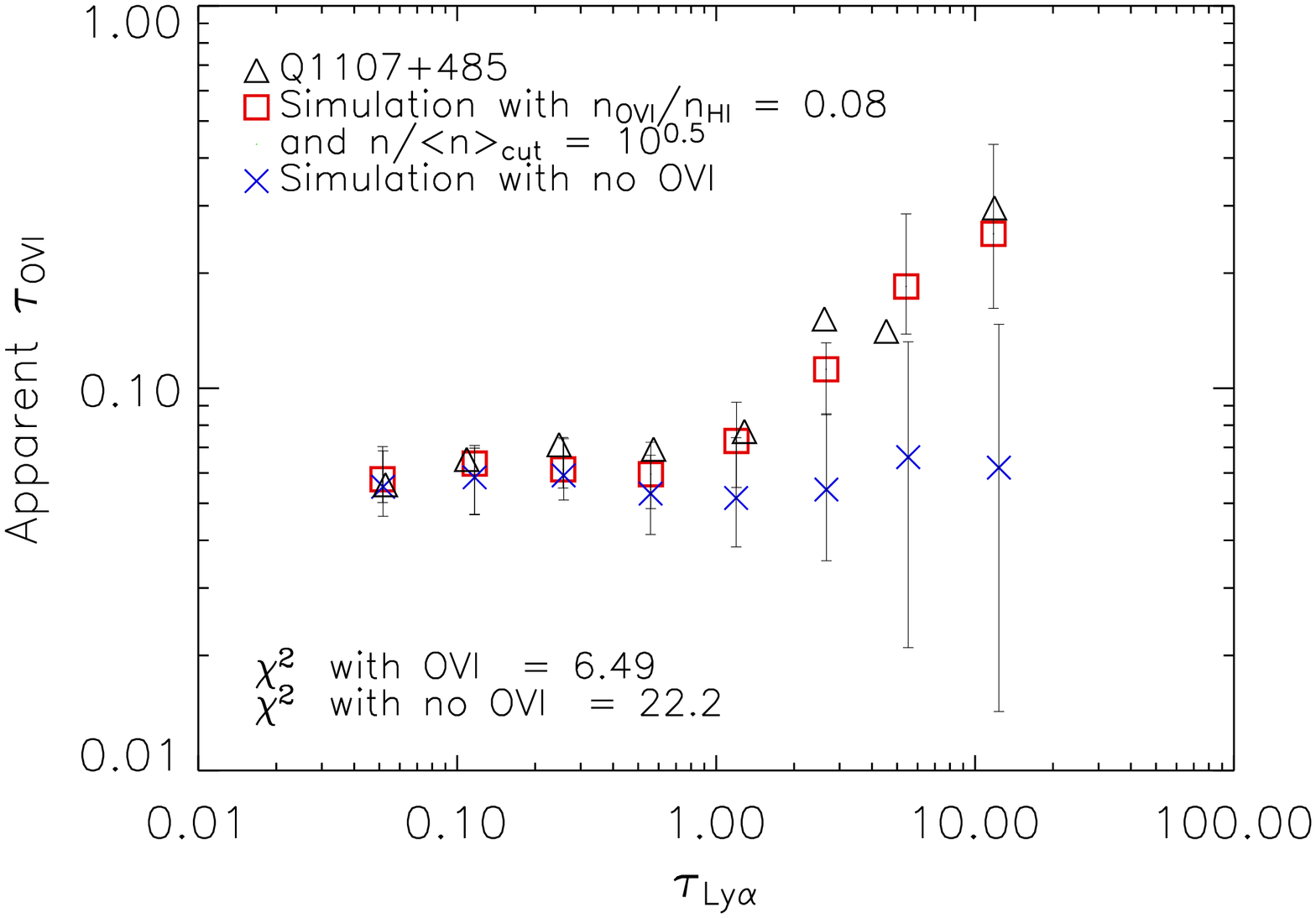}
\includegraphics[angle=90,width=.89\columnwidth]{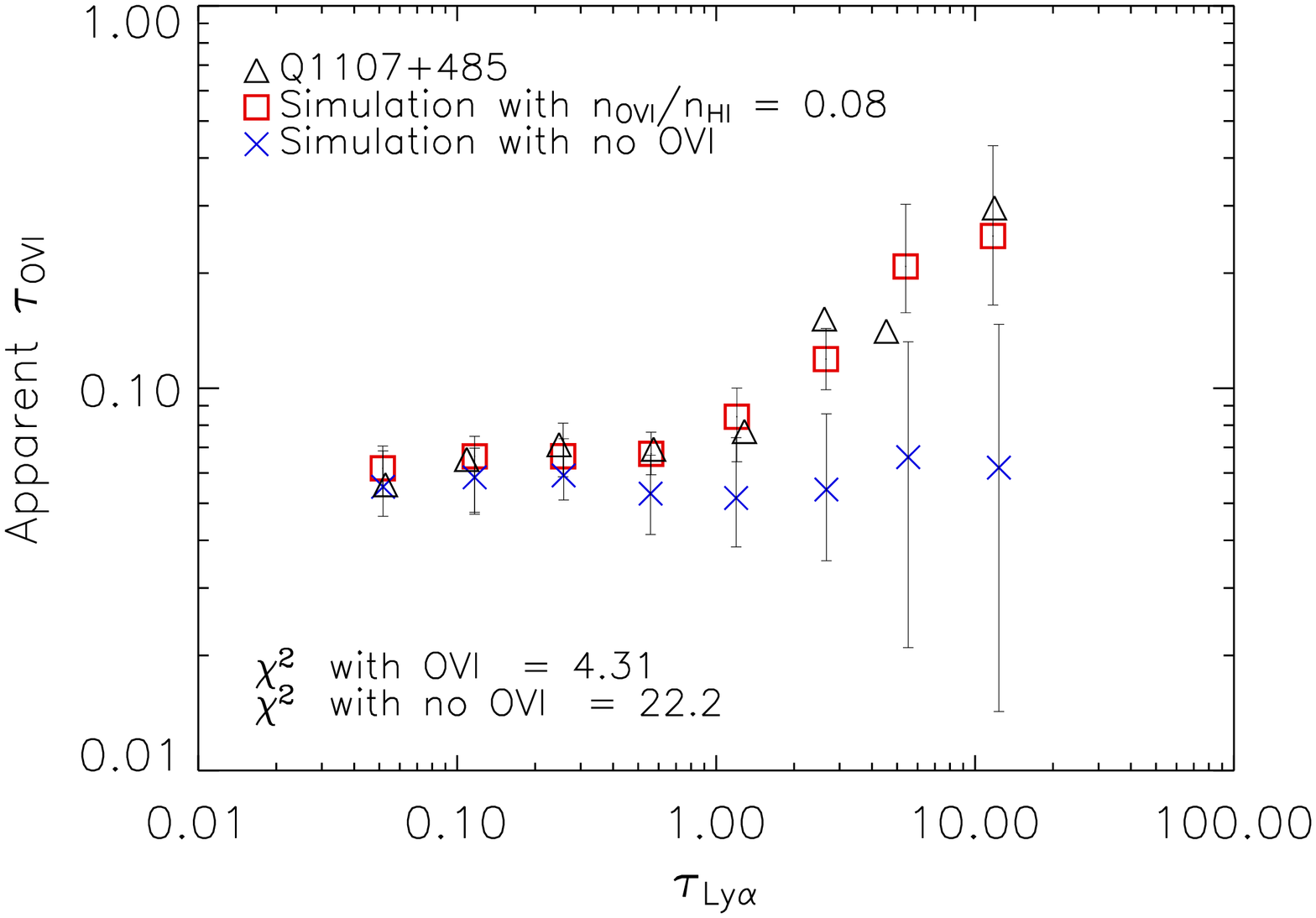}
\includegraphics[angle=90,width=.89\columnwidth]{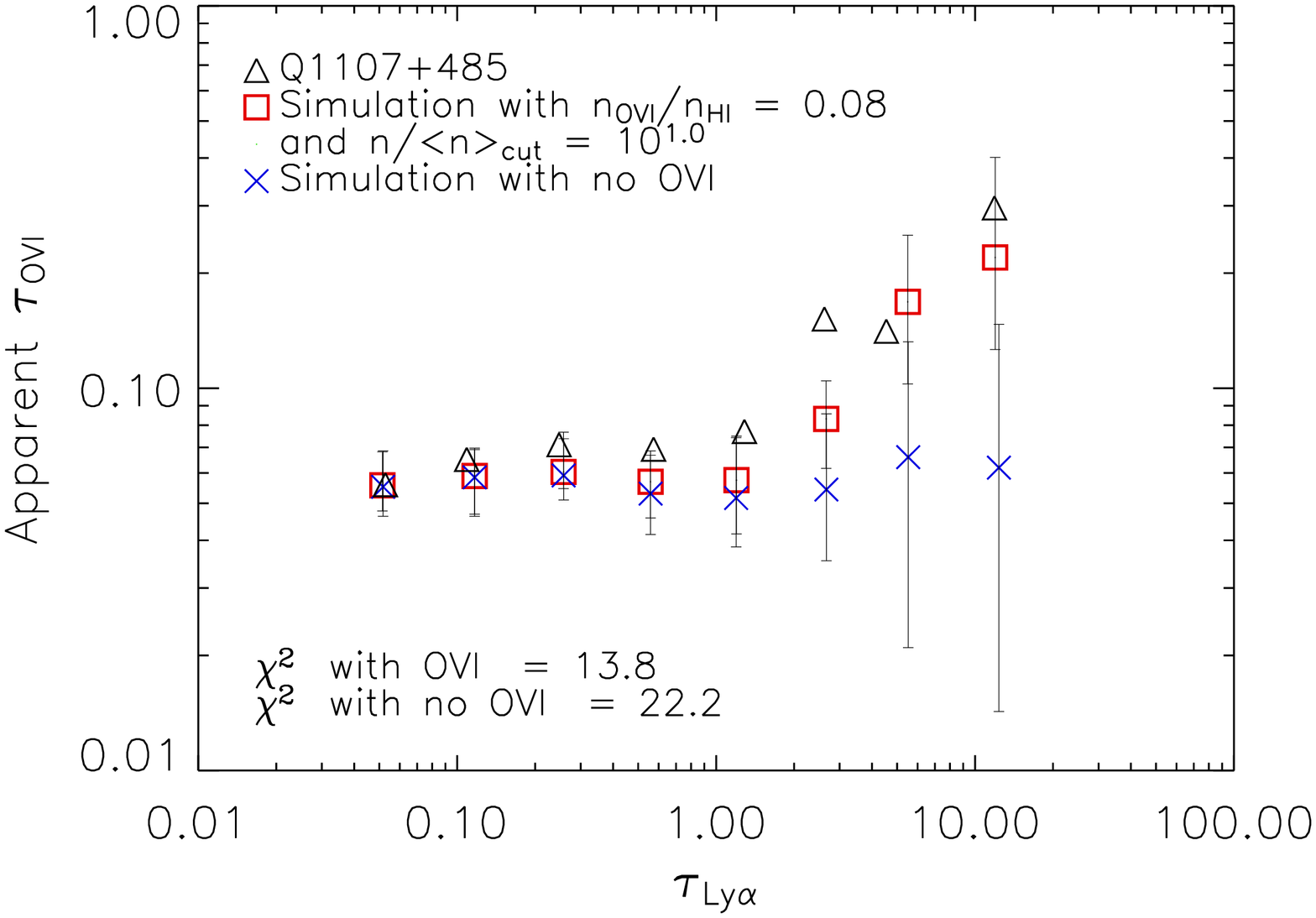}
\includegraphics[angle=90,width=.89\columnwidth]{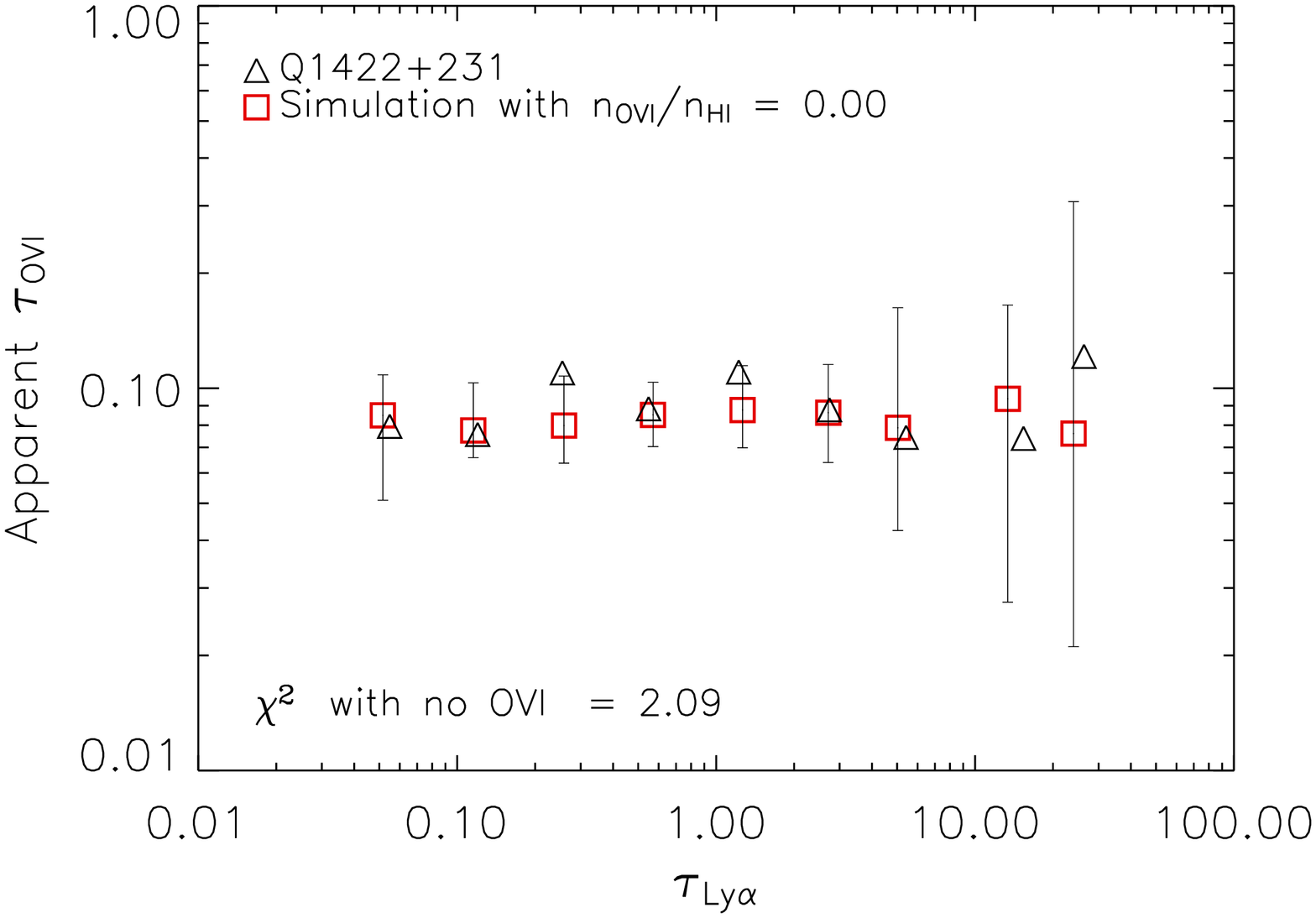}
\includegraphics[angle=90,width=.89\columnwidth]{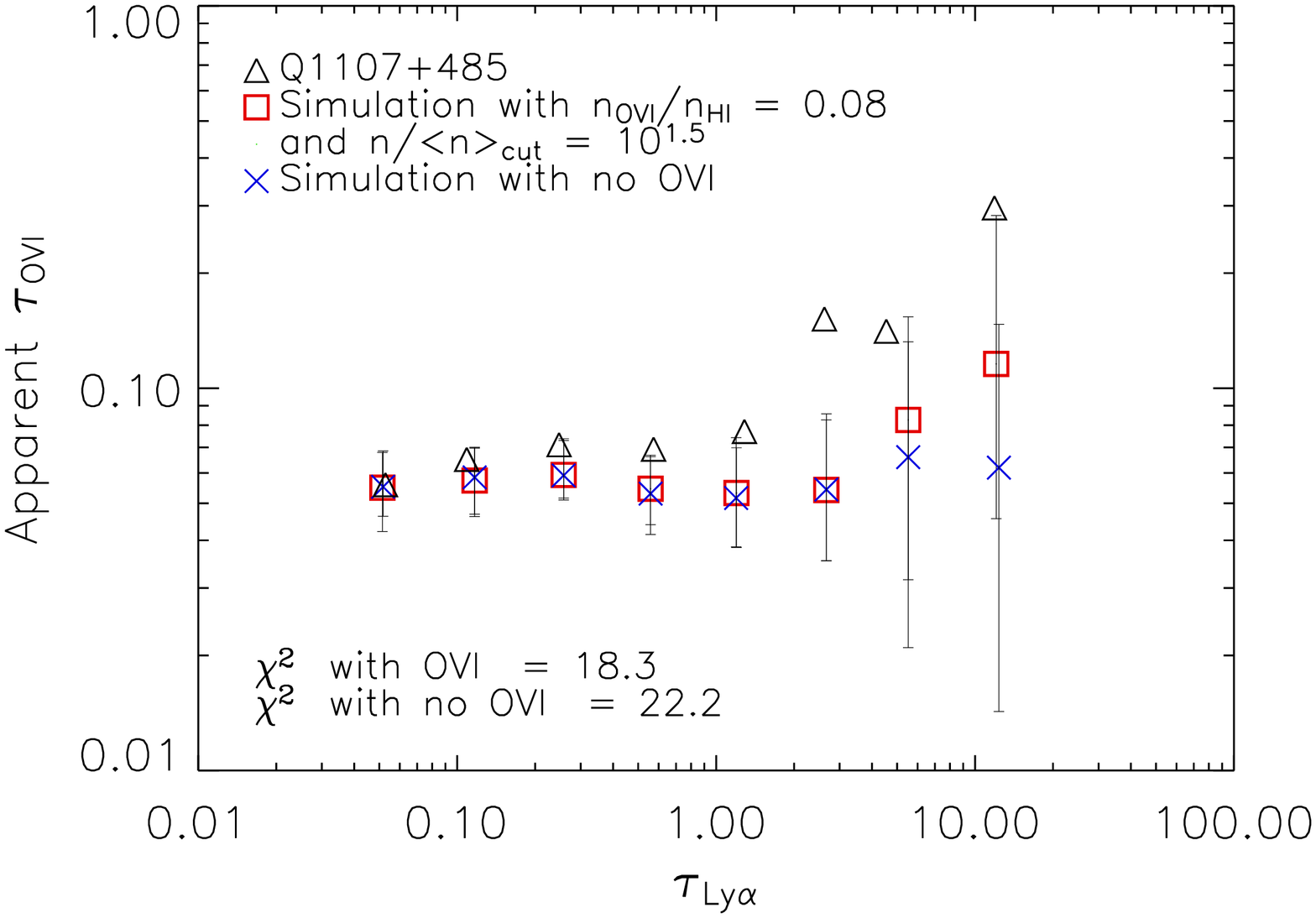}
\caption{{\it Right:} the \taurela relation for observed and simulated synthetic spectra 
for four  QSOs; this is shown for synthetic spectra with both the best 
best fitting constant ratio of $n_{\rm{O\textsc{vi}}}/n_{\rm{H\textsc{i}}}$ and
 with no \ovi. {\it Left:} the effect of varying \overdencut on
synthetic spectra of Q1107 compared to the \taurela relation of the
observed spectrum.}
\label{searchplots}
\end{figure*}

\section{Search for \ovi in the low density IGM}
\label{Search for OVI in the low density IGM}

\subsection{Comparison of real and simulated spectra}
\label{Comparison Of Real And Simulated Spectra}

Here we perform a detailed comparison between four observed QSO
spectra and sets of synthetic spectra. We use the same S/N,
$\bar D_{\rmn{O\textsc{VI}}}$, $\bar D_{\rmn{H\textsc{I}}}$ 
and wavelength
range as in the observed spectra with one exception. In the case of 
Q1122 we decrease $\bar D_{\rmn{O\textsc{VI}}}$ by 35\% in order
to  reproduce the level of apparent $\tau_{\rmn{O\textsc{vi}}}$ at 
low $\tau_{\rmn{H\textsc{i}}}$. The need for this is most likely due 
to a deficit of saturated regions in our synthetic spectra. The 
contribution of saturated regions to the mean flux decrement is high 
at low redshift and the failure to adequately reproduce the high 
incidence of these systems leads to a bias in the mean flux decrement 
\citep{2003astro.ph..8078V}. 

The triangles in the left panel of 
Fig.~\ref{searchplots} show the \taurela relation
for the four observed spectra. For the synthetic spectra we assume
as before that the  \ovi distribution takes the form described in
equations (\ref{delcutrangezero}) and (\ref{delcutrangeconst}). We have 
varied the \ovi density and the overdensity threshold for addition of \ovi 
over the range $n_{\rmn{O\textsc{VI}}}/n_{\rmn{H\textsc{I}}} = 0-0.3$ and 
$(n/\bar n)_{\rmn {cut}} = 0-100$. We have produced samples of
40 synthetic spectra for each set of values. We have then
calculated the \taurela relation  for each sample and assessed the
agreement with the observed relation by calculating  $\chi^2$.
Note, however,  that there may be additional sources of error
such inhomogeneities in the metal distribution which our
Monte Carlo technique does not take into account.
The crosses and squares in the left  panels of  
Fig.~\ref{searchplots} show the \taurela
relation  with no \ovi and the best fitting values for
 $n_{\rmn{O\textsc{VI}}}/n_{\rmn{H\textsc{I}}}$ (\overdencut$=0$),  
respectively.  The best fitting value has a reduced $\chi^2_{\rmn{r}}$ 
close to or somewhat smaller than one. 
There is thus good  agreement between the \taurela  relation  
for our best fitting simulations and the real data. In the right panels
of  Fig.~\ref{searchplots} we use the \taurela relation of 
synthetic and observed spectra for Q1107 to provide an 
example of the difficulty in constraining \overdencut.

\begin{figure}
\centering
\includegraphics[angle=90,width=1.\columnwidth]{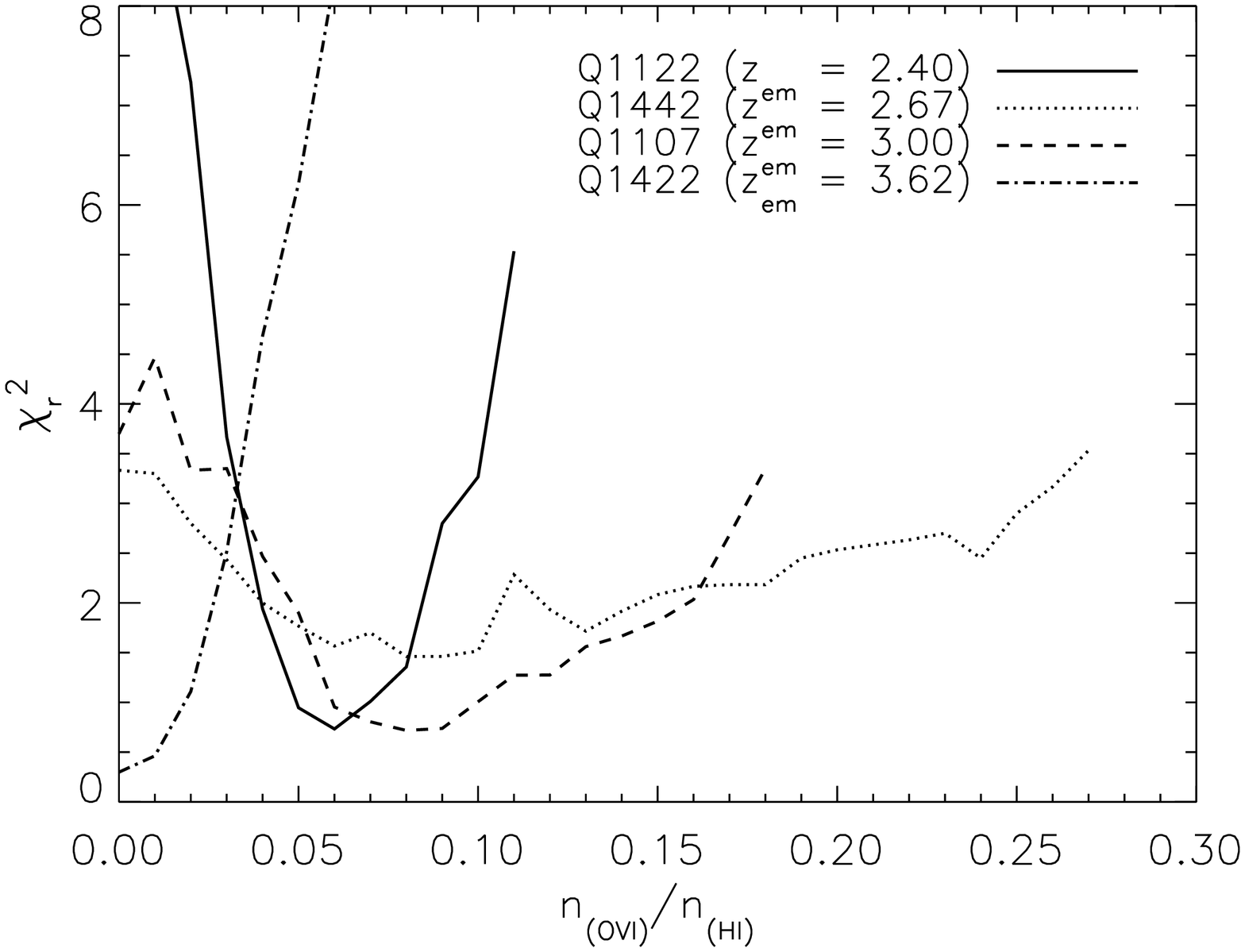}
\includegraphics[angle=90,width=1.\columnwidth]{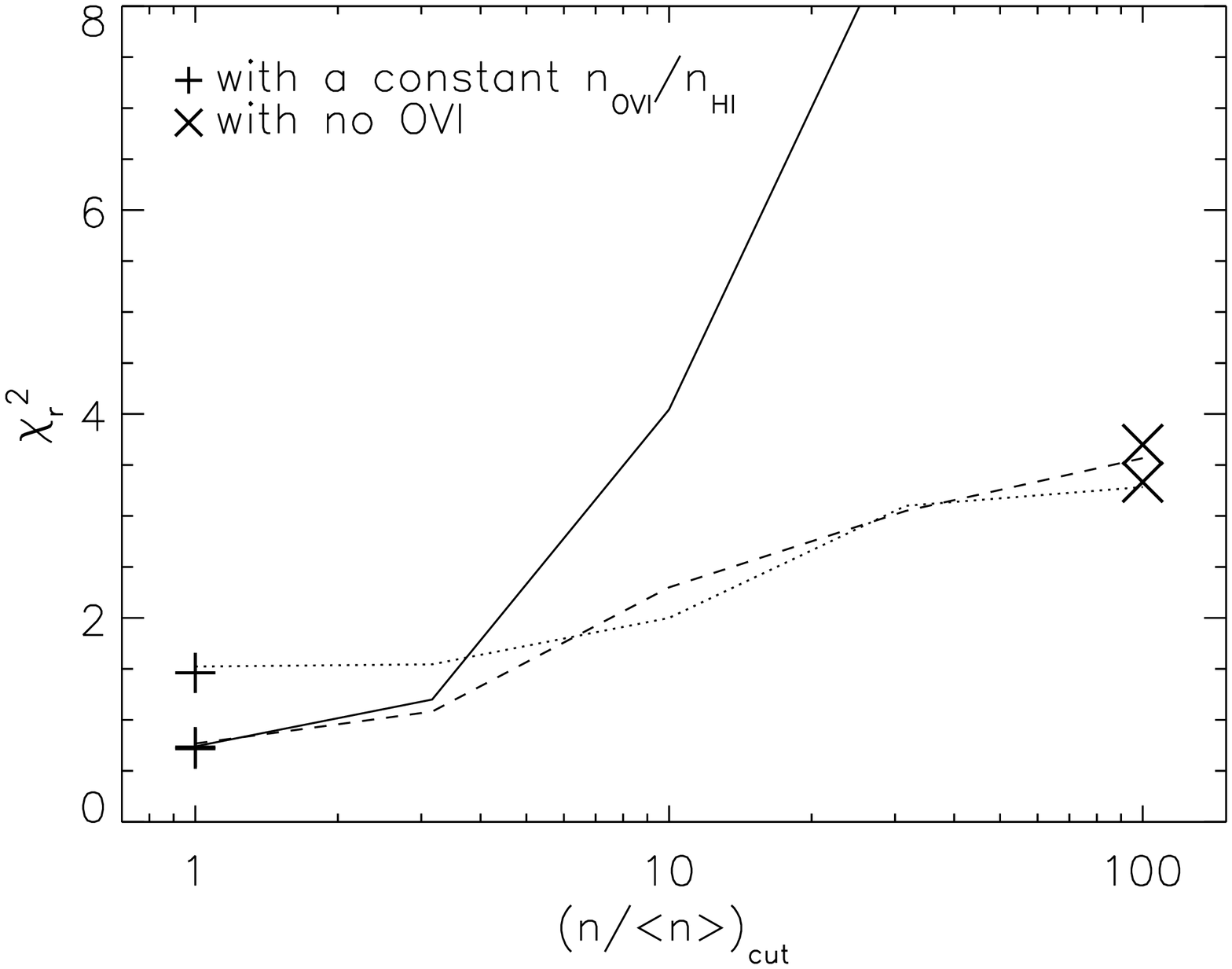}
\caption{Results of a $\chi^2_{\rmn{r}}$ test of the agreement between simulated
and observed $\mathrm{\tau_{O\textsc{vi}}}$. The top panel shows $\chi^2_{\rmn{r}}$ for 
varying $n_{\rmn{O\textsc{vi}}}/n_{\rmn{H\textsc{i}}}$.  The bottom panel shows 
$\chi^2_{\rmn{r}}$ for varying \overdencut. The $\chi^2_{\rmn{r}}$
values for no \ovi 
and a constant $n_{\rmn{O\textsc{vi}}}/n_{\rmn{H\textsc{i}}}$ are also shown.}
\label{rcutchisq}
\end{figure}

The top panel of Fig.~\ref{rcutchisq} shows the reduced $\chi^2_{\rmn{r}}$ as a
function of $n_{\rmn{O\textsc{VI}}}/n_{\rmn{H\textsc{I}}}$ for the
four observed QSOs.  In Q1442 there is a marginal detection of \mbox{O\,{\sc vi}}, 
in Q1107 \ovi is detected with a poorly constrained 
\ovi density  ($n_{\rmn{O\textsc{VI}}}/n_{\rmn{H\textsc{I}}} 
\approx 0.08$)
while in Q1122 \ovi is clearly detected  with 
$n_{\rmn{O\textsc{VI}}}/n_{\rmn{H\textsc{I}}} \approx 0.06$.
 In Q1422 no \ovi is detected. 
Indeed, we find that the spectrum of Q1422 is inconsistent
with \ovi absorption at the same level as detected in Q1122 and Q1107 
with a confidence of greater than 99\%. 

In the bottom panel of Fig.~\ref{rcutchisq} we show how the reduced  
$\chi^2_{\rmn{r}}$
varies as a function of $(n/\bar{n})_{\rmn{cut}}$.We have assumed that 
$n_{\rmn{O\textsc{VI}}}/n_{\rmn{H\textsc{I}}}$ and \overdencut are independent in the 
calculation of the confidence level. \overdencut of greater than 4
 for Q1122, 7 for Q1107 and 4 for Q1442 are ruled out with 95\% confidence. There 
is thus no significant detection 
of \ovi at overdensities $\la 5$.

\subsection{Other searches for \ovi absorption in the
observed sample} 

\citet{2000ApJ...541L...1S} have claimed that  the correlation of 
apparent \ovi
optical depth and \hi extends to $\tau_{\rmn{H\textsc{i}}} \sim  0.1$ 
in the case of Q1122, Q1107 and Q1442. They have interpreted this as 
a detection of \ovi in underdense
regions with \overden$\sim 0.3-0.5$.  We cannot confirm this
with our more detailed analysis.
\citet{1998ApJ...509..661D} found that Q1422 is
consistent with no \ovi in a search for weak \ovi absorption lines
consistent with our result. Note that they detected \ovi in a 
\mbox{O\,{\sc vi}}-\civ pixel correlation search which probed
large \lya optical depth. 
In the spectrum of Q1442, \citet*{2002ApJ...578..737S}
found two strong  \ovi systems associated with high column density
\hi absorption systems but no weak absorption systems 
consistent with our marginal detection. 
\citet*{2002ApJ...578...43C} found a large 
number of \ovi absorption systems associated with low and high column density
\hi absorption systems in the spectrum of Q1122
again consistent with our results.

\subsection{Implications for the metal enrichment of the IGM}
\label{Implications For Metal Enrichment}

\begin{figure} \centering
\includegraphics[angle=90,width=1.\columnwidth]{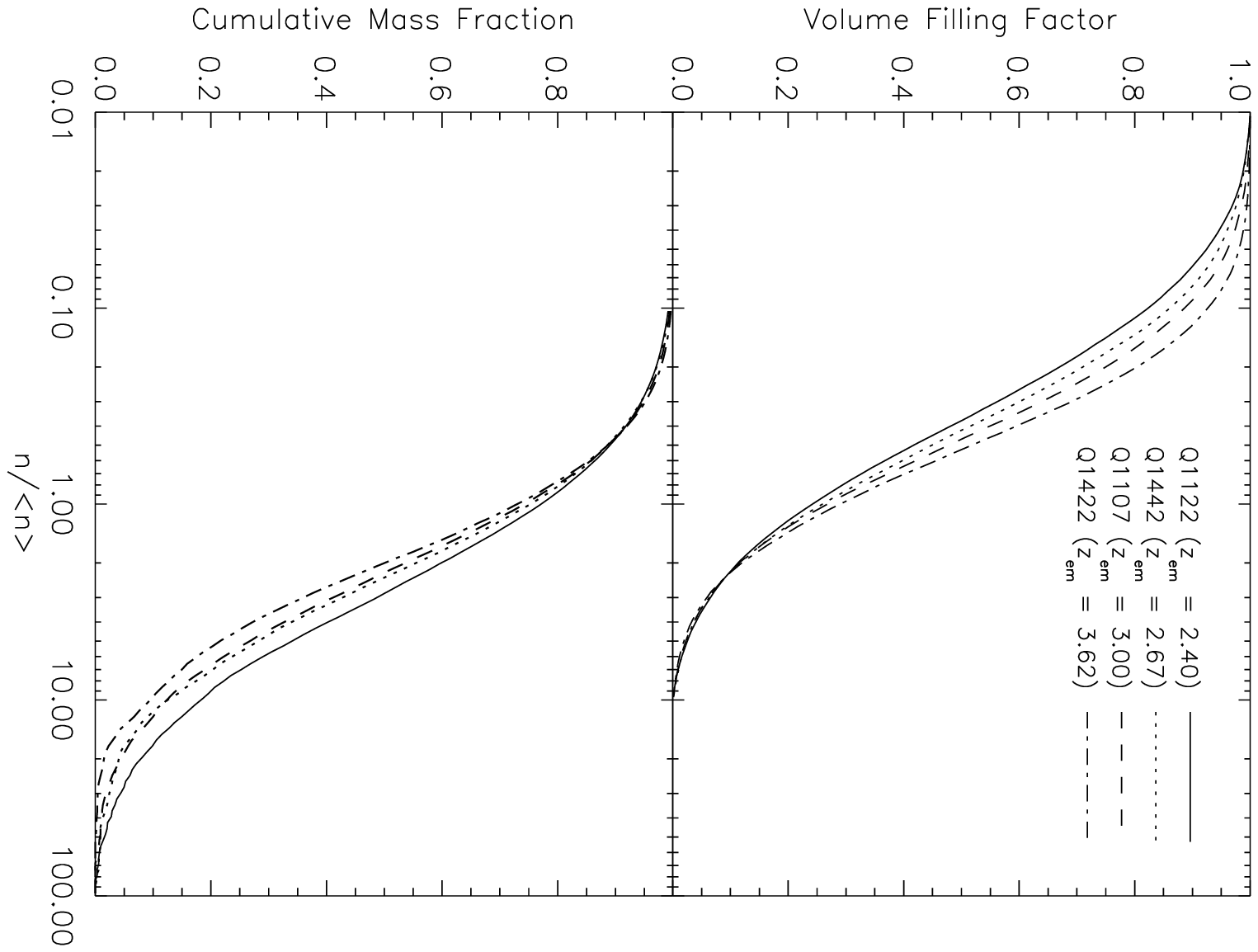}
\caption{The volume filling factor (top panel) and the mass fraction (bottom
panel) as a function of gas overdensity for a lognormal density PDF
at the redshift of the  four observed QSOs.}
\label{filling factor}
\end{figure}

The lack of a significant detection of \ovi at 
low densities cannot be used as evidence 
against the presence of these metals.
As discussed extensively in the previous section, \ovi at these
densities may be masked by \hi absorption. Also, at high redshift the
spectrum of the UV background may not be hard enough to ionise
oxygen up to \mbox{O\,{\sc vi}}. The decreasing contamination by \hi absorption
(due to an expected lower $\mathrm{\bar{\tau}_{Ly\alpha}}$) is
likely to be the main reason why the lowest redshift QSO in
the sample is the only one for which we clearly detect \ovi at
moderate overdensities. 

Fig.~\ref{filling factor} shows how the volume filling factor depends on
overdensity for the lognormal density PDF in the redshift ranges of
the observed QSO absorption. For this PDF the 95\% confidence limits for 
the density threshold,
$(n/\bar{n})_{\rmn{cut}}$, translate into lower 
limits of $4\%$, $1.5\%$ and $4\%$ for the volume 
filling factor of \ovi for Q1122 ($\mathrm{z=2.0-2.3}$), 
Q1107 ($\mathrm{z=2.7-3.0}$) and Q1442 ($\mathrm{z=2.5-2.6}$), respectively. 
There appears to be no evidence
for  a large volume filling factor of metals from the observed
oxygen absorption.  A lower limit of $1.5-4\%$ for the volume filling factor is
no larger than that inferred for winds from Lyman break galaxies
\citep{2003ApJ...584...45A}. A picture where metal enrichment is
due to winds from rather large galaxies at $z\sim 2-5$ is
consistent with the observed \ovi absorption in QSO spectra.
\citet{1998yugf.conf..249H} and \citet{2003astro.ph..5413P} find
that the same is true for the observed \civ absorption, although 
\citet{2003astro.ph..6469S} may find new evidence of relevance.

Fig.~\ref{filling factor} also shows the relation between mass fraction and 
overdensity for the lognormal density distribution. Despite the rather low 
inferred volume filling factor for which metals have been detected, about 
20-40 \%  of baryons must already be enriched with metals 
to explain the observed absorption.

\section{Conclusions}
\label{Conclusions}

We have used a large sample of synthetic spectra
which mimic the observational properties of four
observed QSO spectra to interpret the results of a search
for \ovi in the low density IGM.  Our results can be summarised as follows.

\begin{enumerate}

\renewcommand{\theenumi}{(\arabic{enumi})}
 \item At low \lya optical depth the corresponding  apparent 
  \ovi optical depth obtained  by  using the pixel correlation technique
  developed by \citet{1998Natur.394...44C} and \citet{2000ApJ...541L...1S} 
  depends mainly on the    mean flux decrement in the \ovi region of 
  the spectrum.

\item In a sample of four QSO in the redshift range 2-3.5 we detect
   \ovi significantly  in two  QSOs and marginally in the  third.
   The significance of the detection increases with decreasing
   redshift.

\item The position of the bend in the relation of 
      apparent \ovi and \lya optical depth depends only 
      weakly on the lowest density  at which OVI 
      is present in the IGM.  

\item We obtain upper limits of 4, 7 and 4 for the minimum density
   for which \ovi has been detected with 95\% confidence. For the lognormal model 
   density distribution this translates into
   lower limits of  $4\%$, $1.5\%$ and $4\%$ for the volume 
   filling factor of metals.
   We thus do not confirm previous claims of a detection of \ovi in
   underdense regions and a corresponding large volume filling
   factor of metals.

\item The \ovi absorption in QSO absorption spectra as detected
   by the pixel correlation technique
   provides no evidence for (or against) a widespread metal
   enrichment at very high redshift ($z\sim 10-20$). The lower limit
   for the volume filling  factor of metals is equally
   consistent with  metal enrichment by winds from
   Lyman break galaxies.

\end{enumerate}

\section*{Acknowledgements}

We would like to thank Michael Rauch, Len Cowie and ESO  for providing 
the observed spectra and the referee Anthony Aguiree for a helpful report.  
The authors would further like to thank Steve Warren  for his useful 
suggestions  and  Alex King and Thomas Babbedge for helpful comments 
on the manuscript. This work was supported by the European Community 
Research and Training Network ``The Physics of the Intergalactic 
Medium''

\bibliography{biblio2}
\label{lastpage}

\end{document}